\documentclass[aps,pre,twocolumn,floatfix,groupedaddress,superscriptaddress, showpacs]{revtex4-1} 

\usepackage{amsmath}
\usepackage{amssymb}
\usepackage{lipsum}
\usepackage{mathrsfs}

\usepackage{graphicx}
\usepackage{bm}
\usepackage[T1]{fontenc}
\usepackage{url}
\usepackage[bf]{subfigure}
\usepackage{rotating}
\usepackage{soul}
\usepackage{scalefnt}
\usepackage[hidelinks]{hyperref}
\usepackage[utf8]{inputenc}
\usepackage{enumerate}
\usepackage{float}
\usepackage{epigraph}

\usepackage{amsthm}

\theoremstyle{plain}

\newtheorem{lem}{Lemma}

\usepackage{color}
\definecolor{redcolor}{rgb}{1.0,0.,0.}

\begin{document}
\title{Principal Component Analysis: A \emph{Natural} Approach to Data Exploration}

\author{Felipe L. Gewers}
\affiliation{S\~ao Carlos Institute of Physics, University of S\~ao Paulo, S\~ao Carlos, SP, Brazil}
\author{Gustavo R. Ferreira}
\affiliation{Institute of Mathematics and Statistics, University of S\~ao Paulo, S\~ao Paulo, SP, Brazil}
\author{Henrique F. de Arruda}
\affiliation{Institute of Mathematics and Computer Science, University of S\~ao Paulo, S\~ao Carlos, SP, Brazil}
\author{Filipi N. Silva}
\affiliation{S\~ao Carlos Institute of Physics, University of S\~ao Paulo, S\~ao Carlos, SP, Brazil}
\affiliation{School of Informatics, Computing and Engineering, Indiana University, 
Bloomington, Indiana 47405, USA}
\author{Cesar H. Comin}
\affiliation{Department of Computer Science, Federal University of S\~ao Carlos, S\~ao Carlos, SP, Brazil}
\author{Diego R. Amancio}
\affiliation{Institute of Mathematics and Computer Science, University of S\~ao Paulo, S\~ao Carlos, SP, Brazil}
\affiliation{School of Informatics, Computing and Engineering, Indiana University, 
Bloomington, Indiana 47405, USA}
\author{Luciano da F. Costa}
\affiliation{S\~ao Carlos Institute of Physics, University of S\~ao Paulo, S\~ao Carlos, SP, Brazil}

\begin{abstract}
Principal component analysis (PCA) is often used for analysing
data in the most diverse areas.  In this work, we report an
integrated approach to several theoretical and practical aspects
of PCA.  We start by providing, in an intuitive and accessible
manner, the basic principles underlying PCA and its applications.
Next, we present a systematic, though no exclusive, survey of
some representative works illustrating the potential of PCA
applications to a wide range of areas.  An experimental 
investigation of the ability of PCA for variance explanation
and dimensionality reduction is also developed, which confirms
the efficacy of PCA and also shows that standardizing or not
the original data can have important effects on the obtained
results.  Overall, we believe the several covered issues can 
assist researchers from the most diverse areas in using and
interpreting PCA.
\end{abstract}

\maketitle

\begin{quotation}
\epigraph{``Frustra fit per plura quod potest fieri per pauciora.''}{\textit{William of Occam.}}
\end{quotation}

\tableofcontents

\section{Introduction}
\label{s:introduction}
Science has always relied on the collection, organization and analysis of measurements or data.  A proverbial example that promptly comes to mind is the criticality of Tycho Brahe's measurements for the development of Galileo's gravitation studies~\cite{ferguson2002tycho}.  Since that time, substantial technological advances, in particular in electronics and informatics, have implied an ever increasing accumulation of large amounts of the most varied types of data, extending from eCommerce to Astronomy. Not only have more types of data become available, but traditional measurements in areas such as particle physics are now performed with increased resolution and in substantially larger numbers. Such trends are now aptly known as the \emph{data deluge}~\cite{bell2009beyond}.  However, such vast quantities of data are, by themselves, of no great avail unless means are applied in order to identify the most relevant \emph{information} contained in such repositories, a process known as \emph{data mining}~\cite{hand2007principles}.   Indeed, provided effective means are available for mining, truly valuable information can be extracted.  For instance, it is likely that the information in existing databases would already be enough to allow us to find the cure for several illnesses. The importance of organizing and summarizing data can therefore be hardly exaggerated.

\begin{figure}[!]
  \begin{center}
  \includegraphics[width=0.9\columnwidth]{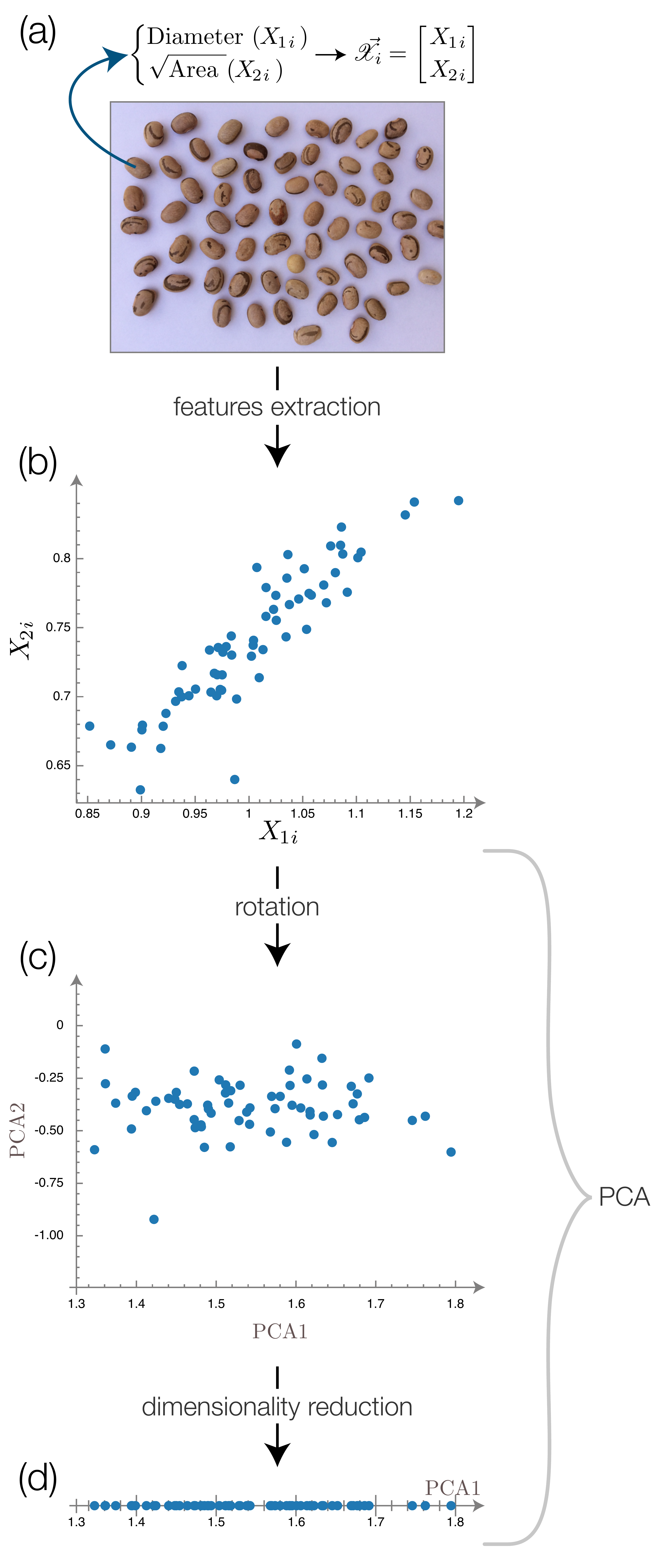} 
  \caption{PCA example on a real-world situation.  Each bean (a) is characterized in terms of two measurements: diameter $X_{1i}$ and square root of area $X_{2i}$. Though these two measurements are intrinsically related in a direct fashion, bean shape variations induce a dispersion of the objects when mapped into the features space (b). PCA allows the identification of the orientation of maximum data dispersion (c).  As the dispersion in the resulting second axis is relatively small, this axis can be discarded (d).
  }
  ~\label{f:PCA_example}
  \end{center}
\end{figure}

While a definitive solution to the problem of data mining remains elusive, there are some well-established approaches which have proven to be useful for organizing and summarizing data~\cite{bishop2006pattern}.  Perhaps the most popular among these, is \emph{Principal Component Analysis -- PCA}~\cite{jolliffe1986principal, da2009shape, abdi2010principal}.  Let's organize the several (N) measurements of each object or individual $i$ in terms of a respective \emph{feature vector} $\vec{\mathscr{X}}_i$, existing in an N-dimensional \emph{feature space}.  PCA can then be understood as a statistical method in which the coordinate axes of the feature space are rotated so that the first axis results with the maximum possible data dispersion (as quantified by the statistical variance), the second axis with the second maximum dispersion, and so on.  This principle is illustrated with respect to a simple situation with $N=2$ in Figure~\ref{f:PCA_example}.  Here, we have $Q$ objects (real beans), each described by $N=2$ respective measurements, which are themselves organized as a respective feature vector.  More specifically, each object $i$ has two respective measurements $X_{1i}$ and $X_{2i}$, giving rise to $\vec{\mathscr{X}}_i$.  In this particular example involving real beans, the two chosen measurements correspond to diameter (i.e. the maximum distance between any two points belonging to the border of the object) and the square root of the bean area.

When mapped into the respective two-dimensional feature space, these objects define a distribution of points which, in the case of this example, assumes an elongated shape.  Finding this type of point distribution in the feature space can be understood as indications of \emph{correlation} between the measurements.  In the case of beans, their shape is not far from a disk, in which the area is given as pi times square radius (equal to half the diameter).  So, except for shape variations, the two chosen measurements are directly related and would be, in principle, redundant.  However, because no two beans have exactly the same shape, we have the dispersion observed in the feature space (Figure~\ref{f:PCA_example}(b)).

The application of PCA to this dataset will rotate the coordinate system, yielding the new axes identified as PCA1 and PCA2 in the figure.  The maximum data dispersion in one dimension is now found along the first PCA axis, PCA1.  The second axis, PCA2, will be characterized by the second largest one-dimensional dispersion.

Interestingly, provided the original data distribution is elongated enough, it is now possible to discard the second axis without great loss of overall data variation.  The resulting feature space now has dimension $M=1$.  

\begin{figure}[]
  \begin{center}
  \includegraphics[width=\linewidth]{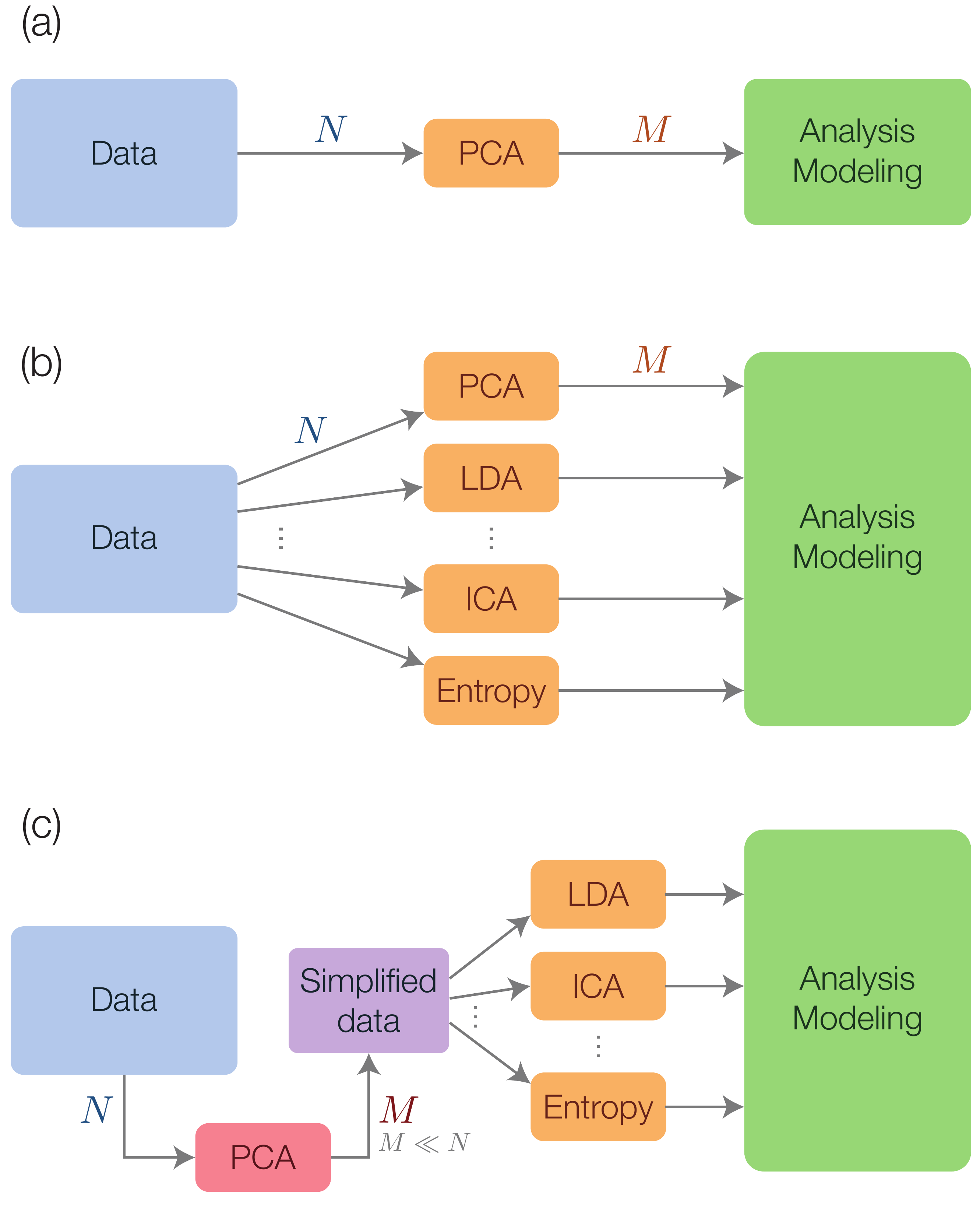} \\
  \caption{The application of PCA in data analysis and modeling. (a) the reference situation, where PCA is applied over the dataset; (b) PCA is applied as one of several complementary data exploration approaches; (c) PCA is used as pre-processing aimed at simplifying the data and improving statistical representation.}
  ~\label{f:PCA_model}
  \end{center}
\end{figure}

The \emph{essence} of PCA applications, therefore, consists in simplifying the original data with minimum loss of overall dispersion, paving the way to a reduction of dimensionality in which the data is represented.  Typical applications of PCA are characterized by having $M << N$.  Observe that PCA ensures maximum dispersion projections and promotes dimensionality reduction, but does not guarantee that the main axes (along to the directions of largest variation in the original data) will necessarily correspond to the directions that would be more useful for each particular study.  For instance, if one is aiming at separating categories of data, the direction of best discrimination may not necessarily correspond to that of maximum dispersion, as provided by PCA.  Indeed, a more robust approach to exploring and modeling a data set should involve, in addition to PCA, the application of several types of projections, including: Linear Discriminant Analysis (LDA), Independent Component Analysis (ICA), maximum entropy, amongst many others~\cite{fodor2002survey, cunningham2008dimension}. This is illustrated in Figure~\ref{f:PCA_model}(a).  However, this approach can imply in substantial computation cost because of the non-linear optimization required by many of the aforementioned projections. In addition, the use of high-dimensional data as input to projection methods can imply in problems of statistical significance~\cite{bishop2006pattern}.  Interestingly, in case of data sets characterized by the presence of correlations between the measurements, PCA can be applied prior to the other computationally more expensive projections in order to obtain data simplification, therefore reducing the overall execution time and catering for more significant statistics. This situation is illustrated in Figure~\ref{f:PCA_model}(b).
Thus, one particularly important issue with PCA regards its efficiency for simplifying, through decorrelation, data sets typically found in the real-world or simulations.  This issue is addressed experimentally in the present work.

To our knowledge, these questions have not been specific and systematically addressed in the context of PCA.  However, extensive evidence exists supporting hypothesis (1), including several examples in which real-world data can have most of its dispersion preserved while keeping just a few of the principal component axis.  However, despite such evidences, it would still be interesting to perform a more systematic investigation of the potential/efficiency of PCA with respect to specific types of data (e.g. biological, astronomical, simulated data, etc.).  

All in all, this work has three main objectives: (a) to present, in intuitive and accessible manner, the concept of PCA as well as several issues regarding its practical applications; (b) to provide a survey of applications of PCA to real-world problems, thus illustrating the potential and versatility of this approach; and (c) to perform an experimental investigation about the ability of PCA to simplify data, through dimensionality reduction, with respect to some of the major areas of knowledge.  In addition, special efforts have been invested to achieve a work that could be interesting to researchers from diverse levels and areas.  For instance, in addition to providing a step-by-step presentation of PCA, more advance issues such as proof of dispersion maximization, stability of the covariance matrix, etc. are provided that will probably be of interest to more experienced readers.

\section{Correlation, Covariance \& Co.}
\label{s:corrCovExp}

The present work assumes datasets organized as in Table~\ref{t:dataOrganiz}, including several objects (or ``individuals''), each characterized by $N$ respective measurements or \emph{features}.  It is important to realize that each of these features actually correspond to a random variable~\cite{bertsekas2002introduction}, which immediately makes explicit the importance of statistics in data analysis.  Each object can be thought as a vector (or point) in an $N$-dimensional \emph{feature space}. Because $N$ is usually larger than 2 or 3, it becomes a challenge to visualize the overall distribution of objects in a typical feature space.  By projecting this space into 2 or 3 dimensions, PCA can be of great help in obtaining \emph{visualizations} of more elaborate datasets.

\begin{table}[]
\begin{tabular}{ccccc}
  \hline
  \hline
			  & Object 1 & Object 2 & \dots & Object Q \\
  \hline
    Feature 1 & 5.4		 & 2.4	    & \dots & 12.3 \\
    Feature 2 & 7.5      & 3.5		& \dots & 10.3 \\
    \vdots	  & \vdots   & \vdots   & $\ddots$ & \vdots \\
    Feature N & 8.3      & 1.4		& \dots & 14.2 \\
    \hline
    \hline
\end{tabular}
\caption{\label{t:dataOrganiz} Typical organization of the input data for PCA application.}
\end{table}

\begin{figure}[!h]
  \begin{center}
  \includegraphics[width=0.6\linewidth]{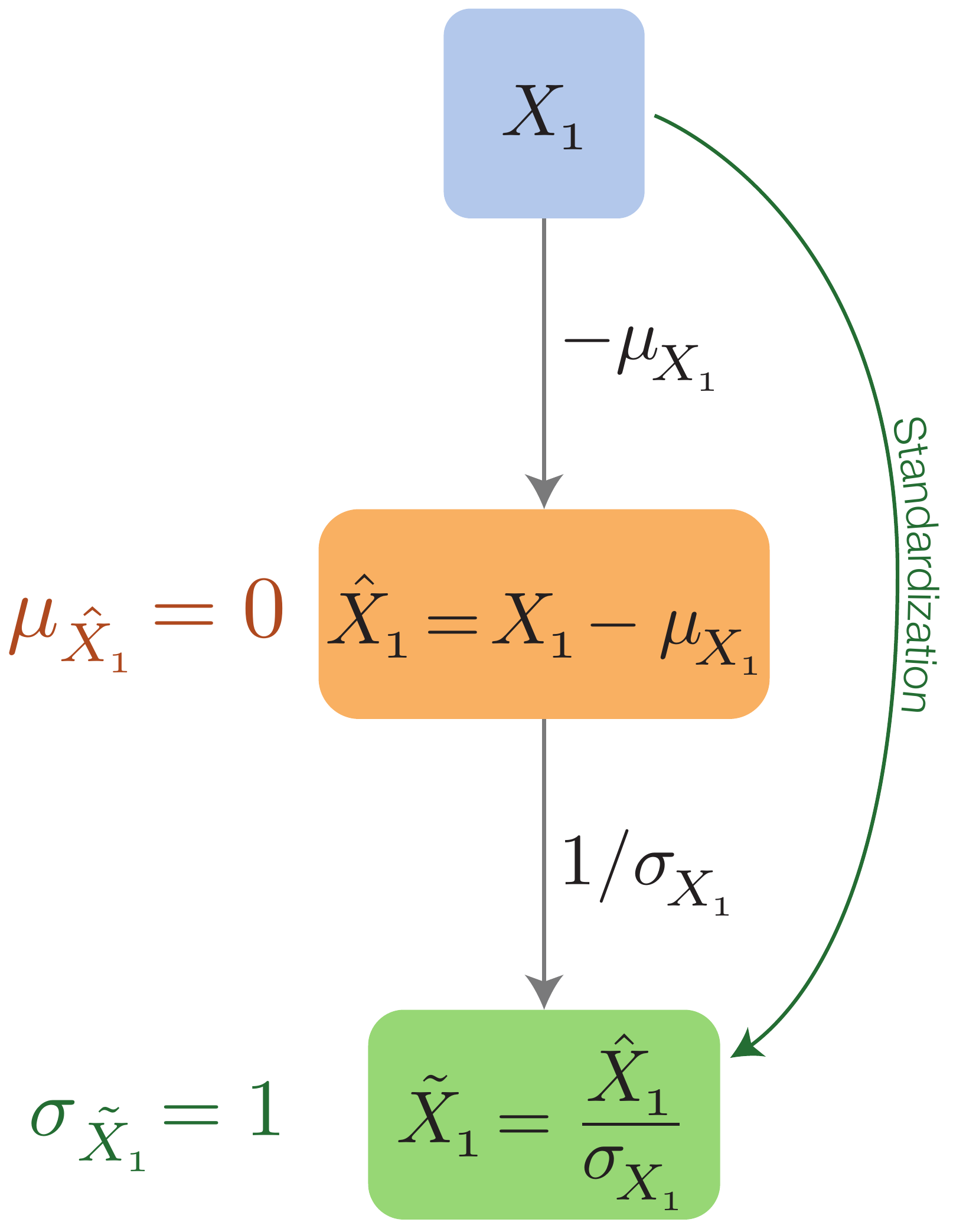} \\
  \caption{An original random variable $X_1$ can be statistically transformed to the coordinates origin by subtracting its respective average, yielding the new random variable $\hat{X}_1$.  A subsequent division by the standard deviation will normalize the variable's dispersion, yielding a dimensionless variable $\tilde{X}_1$.}
  ~\label{f:PCA_stand}
  \end{center}
\end{figure}

Because each object in the original dataset is characterized in terms of $N$ random variables, it becomes immediately possible to implement a series of operations on these objects, such as displacing them to the coordinate origin, normalizing their dispersions, amongst other possibilities. Statistically, such operations can be understood as particular cases of \emph{statistical transformations}~\cite{feller2008introduction}. More specifically, given a random variable $X_1$, any function that maps it into another random variable can be understood as a statistical transformation.  Figure~\ref{f:PCA_stand} illustrates two particularly important such transformations, corresponding to translation to the coordinate origin (by subtracting the respective average $\mu_{X_1}$), and subsequent variance normalization (dividing by the respective standard deviation $\sigma_{X_1}$). The combined application of these two transformations yields the well-known operation of \emph{standardization}~\cite{everitt2002cambridge}.  As often adopted, we will use the term \emph{normalization} to refer to any generic alterations of the original measurements aimed at making them more compatible, reserving the term \emph{standardization} to the specific statistical transformation involving subtraction of the average and subsequent division by the standard deviation.

After translation to the coordinate origin, the new random variable $\hat{X}_1$ will have zero mean. After a random variable $X_1$ is standardized into $\tilde{X}_1$, this new variable will necessarily have zero mean and unit standard deviation (and, thus, unit variance). In addition, most of the observations of this random variable will be comprised in the interval ranging between $-2$ and $2$ due to Chebyshev's inequality~\cite{bertsekas2002introduction}.

\begin{figure}[]
  \begin{center}
  \includegraphics[width=\linewidth]{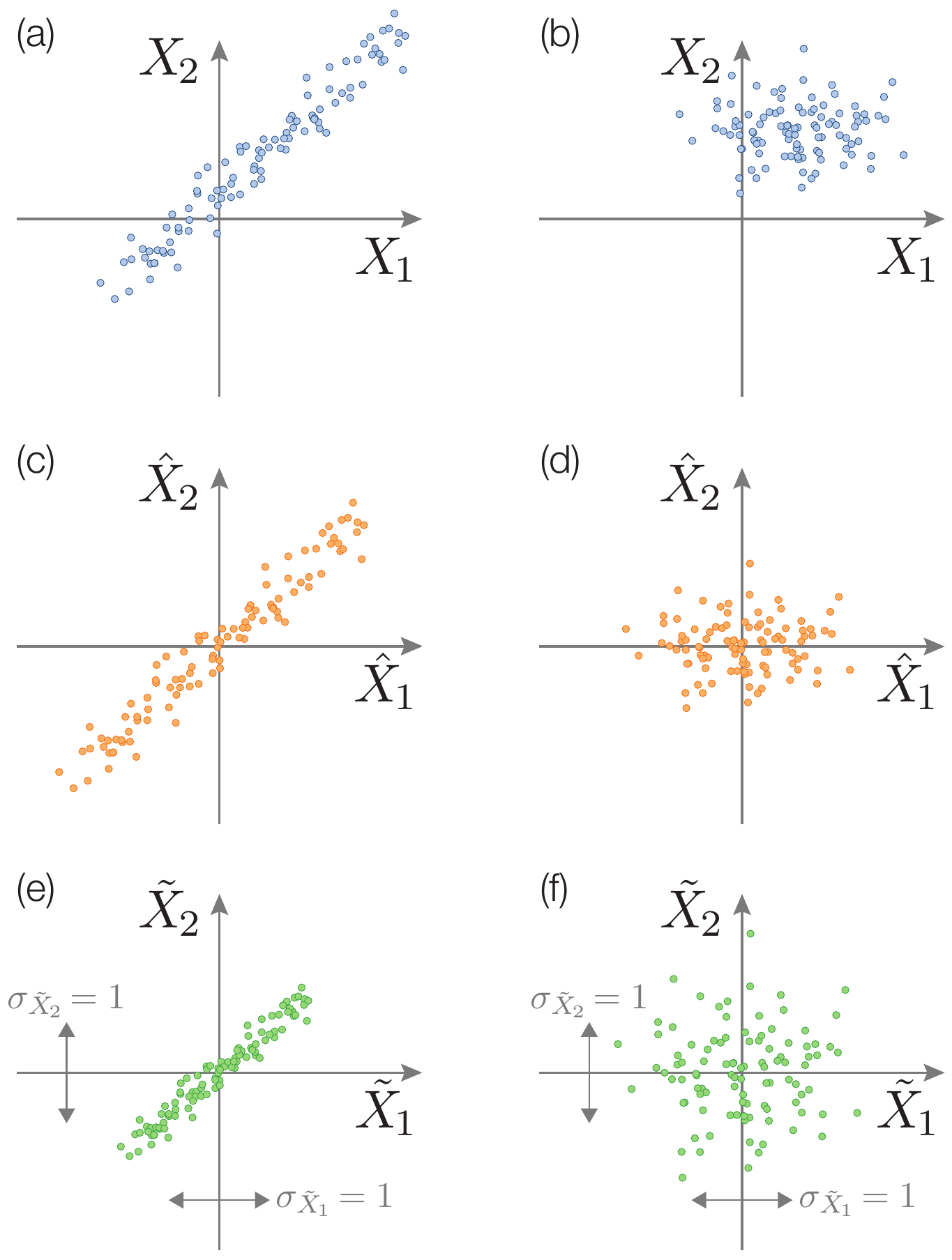} \\
  \caption{Two examples of data distributions (a) and (b) and respective normalizations by average subtraction (c) and (d) and division by standard deviation (e) and (f).}
  ~\label{f:PCA_standExample}
  \end{center}
\end{figure}

Given two random variables $X_1$ and $X_2$, it is important to consider statistical measurements of their possible relationship or joint variation.  Such measurements can then be used to quantify how much two variables are related, an aspect that is directly related to \emph{data redundancy}. There are three main basic ways to do so, as allowed by: \emph{correlation}, \emph{covariance}, and \emph{(Pearson) coefficient of correlation}~\cite{pearson1895note}.  All these three measurements can be conveniently expressed in terms of the expectation of products between $X_1$ and $X_2$. Informally speaking, the expectation $E[X_i]$ of a random variable $X_i$ corresponds to the \emph{average} of that variable. For instance, the \emph{correlation} between $X_1$ and $X_2$ is simply given as:

\begin{equation}
R_{12}=Corr(X_1,X_2)=E[X_1 X_2].
\end{equation}
This quantity already expresses some level of relationship between the two variables. Consider the example in Figure~\ref{f:PCA_standExample}(a), in which $X_1$ has an evident relationship with $X_2$. Most of the products $X_1 X_2$ in this example are positive, implying in positive $E[X_1 X_2]$, identifying a positive correlation value.  Consider now the objects distribution in Figure~\ref{f:PCA_standExample}(b). It can be easily verified that a positive value $E[X_1 X_2]$ will again be obtained, expressing a relationship between $X_1$ and $X_2$, which is indeed true in the sense that \emph{both} these variables tend to have relatively large, positive values.

\begin{figure}[]
  \begin{center}
  \includegraphics[width=\linewidth]{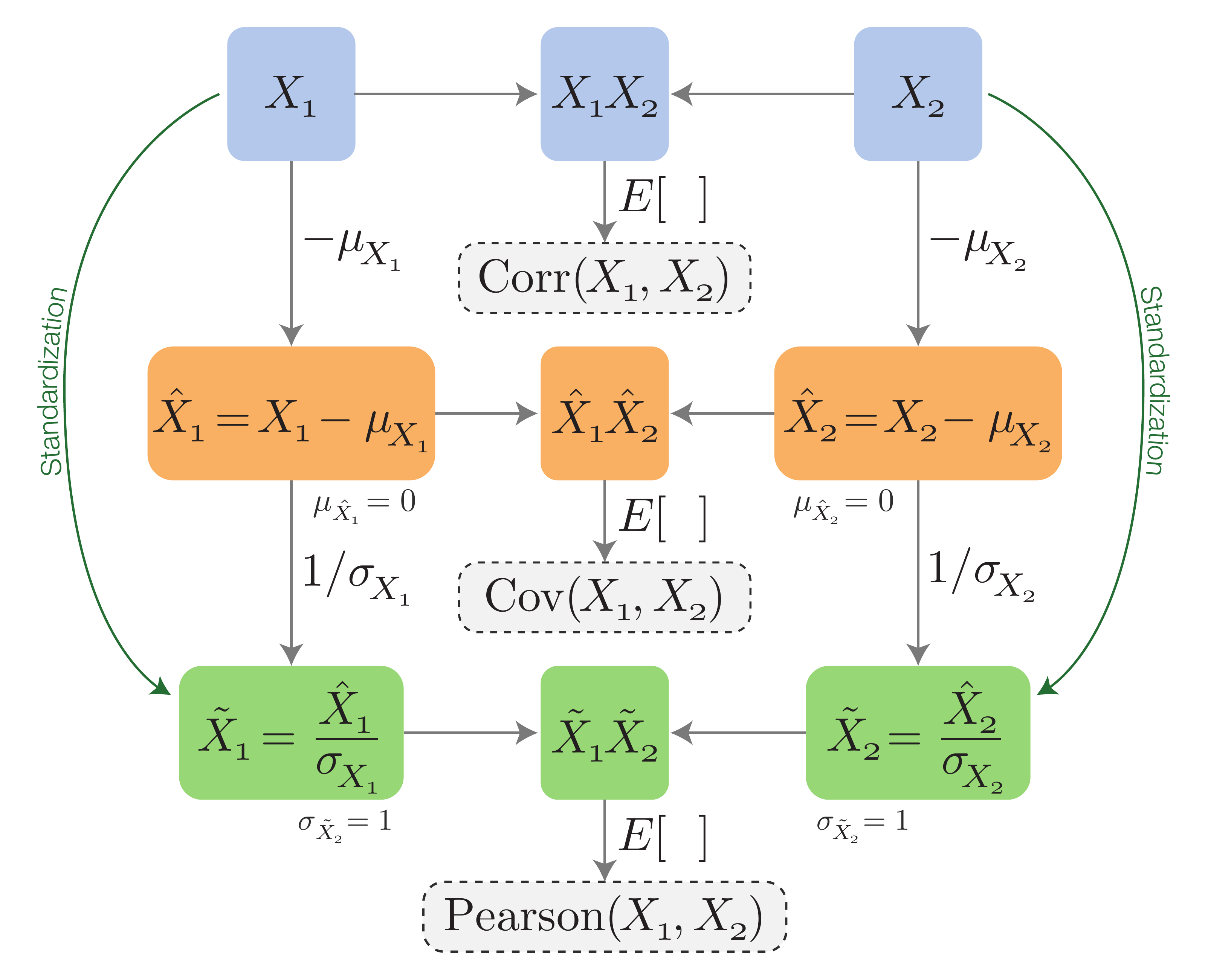} \\
  \caption{The correlation, covariance, and Pearson coefficient of correlation between two given random variables $X_1$ and $X_2$ can be understood as the expectation $E[\;]$ of the products between, respectively: the two original variables, these variables translated to the coordinates origin, and the two original variables moved to the origin and normalized by the respective standard deviations.}
  ~\label{f:PCABasics}
  \end{center}
\end{figure}

Another statistical measurement of relationships between two random variables is the \emph{covariance}. As hinted in its own name, this measurement quantifies joint variations between the two variables. Mathematically, the covariance between $X_1$ and $X_2$ is given as~\cite{everitt2002cambridge}:
\begin{align}
K_{12} = Cov(X_1,X_2) & = Corr(\hat{X}_1,\hat{X}_2) \nonumber\\
					  & =E[(X_1-\mu_{X_1})(X_2-\mu_{X_2})].
\end{align}

Thus, the covariance between $X_1$ and $X_2$ corresponds to the correlation between variables $\hat{X}_1$ and $\hat{X}_2$. Figures~\ref{f:PCA_standExample}(c) and (d) show the effect of moving the original data distributions in Figures~\ref{f:PCA_standExample}(a) and (b), respectively, to the coordinates origin. Observe that the random variables $\hat{X}_1=X_1-\mu_{X_1}$ and $\hat{X}_2=X_2-\mu_{X_2}$ have an expected value of zero. The covariance between $X_1$ and $X_2$ can be estimated as the average of the products $\hat{X_1} \hat{X_2}$, which are all positive in the case of Figure~\ref{f:PCA_standExample}(c), indicating clearly that $X_1$ and $X_2$ present a joint tendency to vary together. However, in the case of the points distribution in Figure~\ref{f:PCA_standExample}(d), the products $\hat{X}_1 \hat{X}_2$ will tend to cancel between the positive values obtained for the quadrants 1 and 3 and the negative values in the quadrants 2 and 4, resulting in nearly null overall covariance between $X_1$ and $X_2$. Observe that the point distribution in Figure~\ref{f:PCA_standExample}(b) therefore yields positive correlation, but nearly null covariance, while positive correlation and covariance are obtained for the points in Figure~\ref{f:PCA_standExample}(a). When two variables $X_1$ and $X_2$ have a null covariance value, they are said to be \emph{uncorrelated}.  

Now we proceed to the (Pearson) coefficient of correlation between $X_1$ and $X_2$. This quantity is defined as:

\begin{align}
C_{12} & = PCorr(\tilde{X}_1,\tilde{X}_2) \\
		   & = Cov(\hat{X}_1/ \sigma_{X_1},\hat{X}_2 / \sigma_{X_2}) \\
           & = E \left[\frac{(X_1-\mu_{X_1})(X_2-\mu_{X_2})}{\sigma_{X_1} \sigma_{X_2}} \right].
\end{align}

The coefficient of correlation between $X_1$ and $X_2$ therefore corresponds to the correlation between $\tilde{X}_1$ and $\tilde{X}_2$, or the covariance between $\hat{X}_1/ \sigma_{X_1}$ and $\hat{X}_2/ \sigma_{X_2}$.  Figures~\ref{f:PCA_standExample}(e) and (f) depict the distributions of points in Figure~\ref{f:PCA_standExample}(a) and (b) after translation to the coordinates origin and division by the variables standard deviations.  The yielding standardized variables $\tilde{X}_1$ and $\tilde{X}_2$  are dimensionless and both have unit variance and standard deviation. This implies the orientation of the main elongation in Figure~\ref{f:PCA_standExample}(a) to change.  The Pearson coefficients of correlation for the point distributions in Figure~\ref{f:PCA_standExample}(e) and (f) can be immediately estimated in terms of the average of the products $\tilde{X}_1,\tilde{X}_2$, which are positive for Figure~\ref{f:PCA_standExample}(e) and nearly null for Figure~\ref{f:PCA_standExample}(f).  It can be shown that $-1 \leq C_{12} \leq 1$ for any situation.  In case $C_{12}=1$ or $C_{12}=-1$, the two random variables are perfectly related by a straight line and are, consequently, totally redundant one another. Indeed, the nearer the absolute value of $C_{12}$ is to one, the more redundant one of the variables is with the other.  The two variables will also be redundant for \emph{relatively} larger values of $Cov{(X_1,X_2)}$, but in a non-normalized way.

In a problem involving $N \geq 1$ random variables, the variables can be organized as a \emph{random vector}, and the mean vector $\vec{\mu}_{X}$ can be calculated by taking the average of each variable independently.  For $N > 1$, it is possible to calculate the correlation, covariance or Pearson correlation coefficient for all pairs of random variables. 

It should be observed that the three statistical joint measurements discussed in this section assume \emph{linear} relationship between pairs of random variables.  Other measurements can be used to characterize non-linear relationships, such as the \emph{Spearman's rank correlation} and \emph{mutual information}~\cite{hair1998multivariate}.


\section{Principal Component Analysis}
\label{s:PCAExp}

In this section, we present the mathematical formulation of PCA. For simplicity's sake, we develop this formulation by integrating the conceptual framework presented in the introduction with the basic statistical concepts covered in Section~\ref{s:corrCovExp}. Consider that the original dataset to be analyzed is given as a  matrix $X$, where each of the $Q$ columns represents an object/individual, and each row $i$, $1 \le i \le N$, expresses a respective measurement/feature $\vec{X}_i$. Also, the values measured for the $j$th object are represented as $\vec{\mathscr{X}}_j$

\begin{figure*}[]
  \begin{center}
  \includegraphics[width=0.8\linewidth]{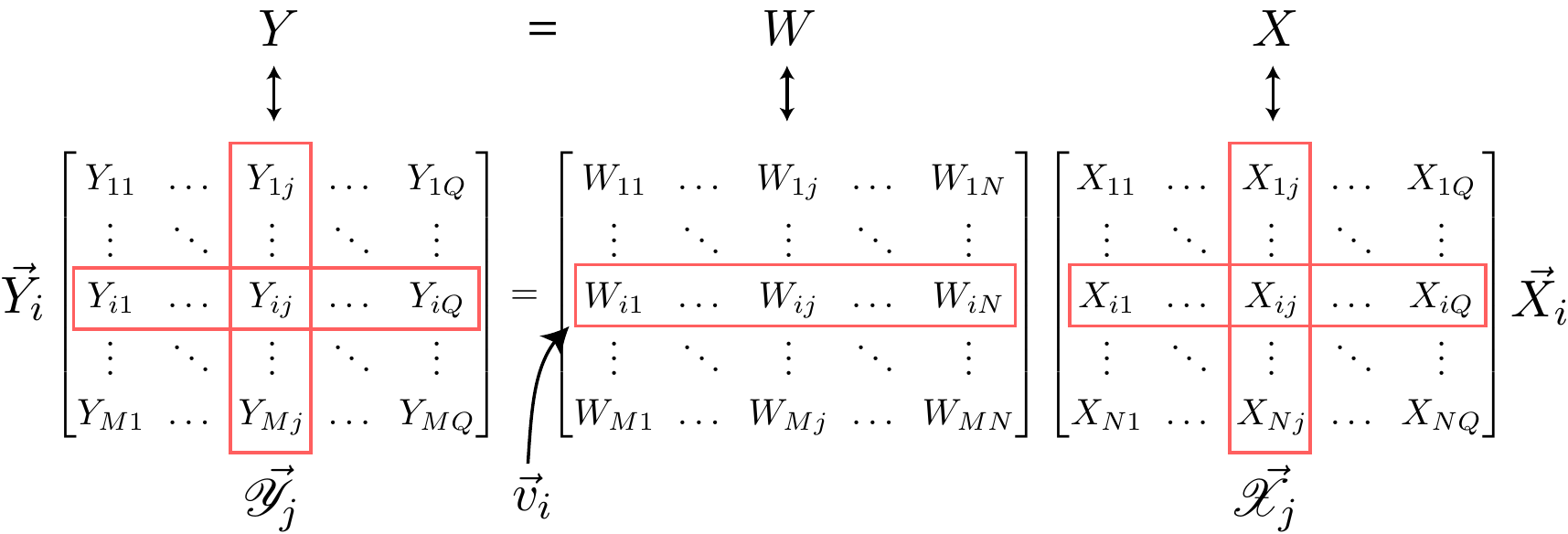} \\
  \caption{The basic components in the PCA transformation (Equation~\ref{eq:PCAProj}). The original data is organized as matrix $X$, each line of which corresponds to a feature vector $\vec{X}_i$. Also, each object is characterized by a vector $\vec{\mathscr{X}}_i$. The transformation matrix, $W$, has each row corresponding to an eigenvector $\vec{v}_i$ of the data covariance or correlation matrix. The $i$th row of the resulting matrix $Y$ contains the $i$th PCA feature, represented as $\vec{Y}_i$, and each projected object $i$ is characterized by the vector $\vec{\mathscr{Y}}_i$.}
  ~\label{f:notation}
  \end{center}
\end{figure*}

An important fact that needs to be taken into account when working with PCA is that, by being a linear transformation, it can be expressed as the following simple matrix form:

\begin{equation}
  Y = W X\label{eq:PCAProj}.
\end{equation}
In other words, the PCA transformation corresponds to multiplying a respective transformation matrix $W$ by the original measurements $X$. Figure~\ref{f:notation} shows the notation used for representing each variable involved in the transformation. All we need to do in order to implement the PCA of a given data is to obtain the respective transformation matrix $W$ and then apply Equation~\ref{eq:PCAProj}.  The derivation of $W$ is explained as follows.

The $i$th element of the empirical mean vector $\vec{\mu}_X = (\mu_{X_1},\dots,\mu_{X_N})^T$, with dimension $N\times 1$, is defined as:

\begin{equation}
\mu_{X_i}=\frac{1}{Q} \sum_{j=1}^Q X_{ij}, 
\end{equation}
where $X_{ij}$ corresponds to the value of the $i$th measurement taken for the $j$th object. The measurements are then brought to the coordinates origin by subtracting the respective means, i.e.:
\begin{equation}
\hat{X}_i = X_i -\mu_{X_i}.
\end{equation}
The matrix containing all the elements $\hat{X}_i$ is henceforth represented as $\hat{X}$. The covariance matrix of the variables in matrix $\hat{X}$ can now be defined as:
\begin{equation}
K = Cov(\hat{X}) = \frac{1}{Q-1}\hat{X}{\hat{X}}^T.
\end{equation}
At this stage we have the covariance matrix of the original measurements. The next step consists in obtaining the necessarily non-negative eigenvalues $\lambda_i$, sorted in decreasing, and respective eigenvectors $\vec{v}_i$, $1 \le i \le N$, of K.

The eigenvectors are now stacked in order to obtain the transformation matrix, i.e.:
\begin{align}
	W =
  	\begin{bmatrix}
    \leftarrow & \vec{v}_1 & \rightarrow \\
    & \vdots & \\
    \leftarrow & \vec{v}_N & \rightarrow \\
  	\end{bmatrix}.
\end{align}
So, all we need to do now to obtain the PCA projection of a given individual $i$ is to use Equation~\ref{eq:PCAProj}, i.e.:

\begin{equation}
\vec{\mathscr{Y}}_i = W \vec{\mathscr{X}}_i,\label{eq:PCAProj_i}
\end{equation}
where $\vec{\mathscr{X}}_i$ and $\vec{\mathscr{Y}}_i$ represent, respectively, the feature vector of object $i$ in the original and projected space.

An important point to be kept in mind is that each dataset will yield a respective transformation matrix $W$.  In other words, this matrix \emph{adapts} to the data in order to provide some critically important properties of PCA, such as the ability to completely decorrelate the original variables and to concentrate variation in the first PCA axes.

So far, the transformed data matrix $Y$ still have the same size as the original data matrix $X$. That is, the transformation implied by Equation~\ref{eq:PCAProj} only remapped the data into a new feature space defined by the eigenvectors of the covariance matrix. This process can be understood as a rotation of the coordinate system that aligns the axes along the directions of largest data variation. Reducing the number of variables corresponds to keeping the first $M \leq N$ PCA axes. The \emph{key question} here is:  what are the conditions allowing this data simplification?  In addition, are there subsidies for choosing a reasonable value for $M$?

The first important fact to consider is that each eigenvalue $\lambda_i$, $1 \leq i \leq N$, of the data covariance matrix $K$ corresponds to the \emph{variance} $\sigma^2_{Y_i}$ of the respective transformed variable $Y_i$.  Let's represent the sum of all these variances as:
\begin{equation}
  S = \sum_{i=1}^{N} \sigma^2_{Y_i}.
\end{equation}

An important property, demonstrated in Section~\ref{s:decorrDemonst}, is that the total data variance $S$ is preserved under axes rotation, and therefore also by the PCA. In other words, the total variance of the original data is equal to that of the new data produced by PCA.  

We can define the \emph{conserved variance} in a PCA with $M$ axes as:
\begin{equation}
  S_c = \sum_{i=1}^{M} \sigma^2_{Y_i}.
\end{equation}
So, the overall conservation of variance by PCA can be expressed in terms of the ratio
\begin{equation}
  G = (100\%)\, \frac{S_c}{S}.\label{eq:PCA_var_ratio}
\end{equation}
Now, the number $M$ of variables to preserve can be defined with respect to G.   For instance, if we desire to preserve $70\%$ of the overall variance after PCA, we choose $M$ so that $G \approx 70\%$.

Many distinct methods have been defined to assist on the choice of a suitable $M$. For instance, Tipping and Bishop~\cite{tipping1999probabilistic} defined a probabilistic version of PCA based on a latent variable model, which allowed the definition of an effective dimensionality of the dataset using a Bayesian treatment of PCA~\cite{bishop1999bayesian}.  In~\cite{hansen1999generalizable}, a generalization error was employed to select the number of principal components, which was evaluated analytically and empirically. In order to compute this error analytically, the authors modeled the data using a multivariate normal distribution. 

In principle, there is no assurance that an $M < N$ exists ensuring that a given variance preservation can be achieved.  This will critically depend on the distribution of the values $\lambda_i$ which, itself, depend on each specific dataset.  More specifically, datasets with highly correlated variables will favor variance preservation. It has been empirically verified that substantial variance preservation can be obtained for many types of real-world data.  Indeed, one of the objectives of the current work is to investigate typical variance preservations that are commonly achieved for several categories of real-world data.

\section{To Standardize or Not to Standardize?}

We have already seen that random variables can be normalized, through statistical transformations, in several ways so as to address specific requirements.  The application of PCA often implies the question whether to normalize or not normalize the original data. Quite often, the dataset is \emph{standardized} prior to PCA~\cite{jolliffe1986principal}, but other normalizations can also be considered.  In this section, we discuss the important issue regarding data standardization prior to PCA.

A possible way to address this issue is to first consider the respective implications.  As seen in Section~\ref{s:corrCovExp}, data standardization of a random variable (or vector) leads to respective dimensionless new variables that have zero mean and unit standard deviation (i.e.~similar scales). Therefore, \emph{all} standardized, dimensionless variables will have \emph{similar} ranges of variation. So, standardization can be particularly advisable as a way to avoid biasing the influence of certain variables when the original variables have significantly different dispersions or scales.  When the original measurements already have similar dispersions, standardization has little effect.

\begin{figure}[htb]
  \begin{center}
  \includegraphics[width=\linewidth]{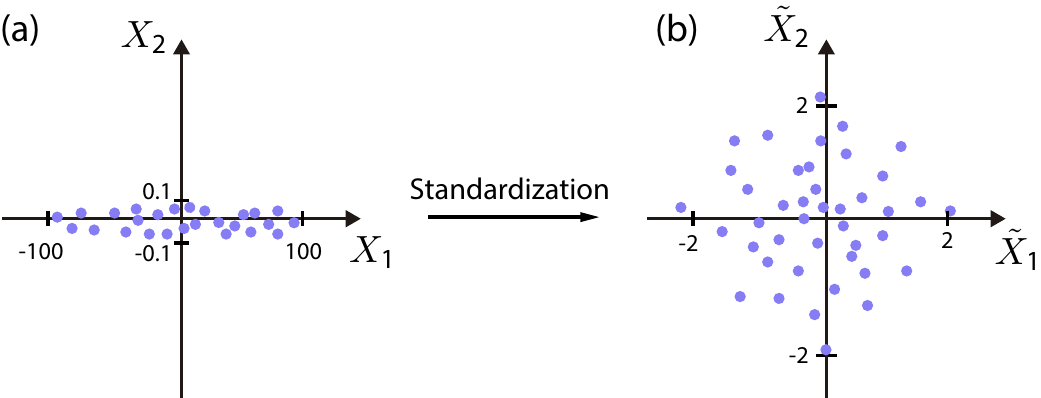} \\
  \caption{A practical situation where standardization should be avoided.  This dataset has a well-defined variation along the first axis, but the dispersion in the second axis is only artifact/noise.  If standardized, the unwanted variation in the second axis will be substantially magnified, therefore affecting the overall 
analysis.}
  ~\label{f:standProblem}
  \end{center}
\end{figure}

There are, however, some situations in which standardization may not be advisable.  Figure~\ref{f:standProblem}(a) shows such a situation, in which one of the variables, namely $X_1$, varies within the range $[-100,100]$, but the other variable, $X_2$, is almost constant other than by a small variation.  In case this small variation is intrinsic to the data (i.e. it is not an artifact) and meaningful, standardization can be used to amplify this information.  However, if this variation is a consequence of an unwanted effect (e.g. experimental error or noise), standardization will emphasize what should have been otherwise eliminated (Figure~\ref{f:standProblem}(b)).  In such cases, either the noise should be reduced by some means, or standardization avoided.  

In order to better understand the influence of standardization on PCA, let's consider two properties $P_1$ and $P_2$ that can be used to characterize a set of objects. Suppose that these two properties are perfectly correlated, that is, their Pearson correlation coefficient is $\rho_{P_1P_2}=1$. When property $P_2$ is measured, an intrinsic error might be incorporated into the measurement. This error may be due to, for instance, the finite resolution of the measurement apparatus or the influence of other variables that were not accounted for in the measurement process. Therefore, the actual measured value $X_2$ may be written as 
\begin{equation}
X_2=P_2+\epsilon, 
\end{equation}
where $\epsilon$ is an additive noise. Suppose that $\epsilon$ is a random variable having normal distribution with mean 0 and variance $\sigma_e^2$. Also, for simplicity's sake, consider that there is no noise associated with the measurement of the other variable $P_1$, that is, $X_1=P_1$. The Pearson correlation coefficient between variables $X_1$ and $X_2$ is given by
\begin{equation}
\rho_{X_1X_2} =\frac{E\left[(X_1-\mu_{X_1})(X_2-\mu_{X_2})\right]}{\sigma_{X_1}\sigma_{X_2}}.
\end{equation}
The mean of the measured variable $X_2$ is $\mu_{X_2}=\mu_{P_2}$, since the noise has zero mean. Supposing that $X_2$ and $P_2$ are normally distributed, the variance of $X_2$ is given by the sum of variances of $P_2$ and $\epsilon$, that is, $\sigma_{X_2}^2=\sigma_{P_2}^2+\sigma_{\epsilon}^2$. Therefore, the Pearson correlation can be expressed as
\begin{align}
\rho_{X_1X_2} & = \frac{E\left[(P_1-\mu_{P_1})(P_2+\epsilon-\mu_{P_2})\right]}{\sigma_{P_1}\sqrt{\sigma_{P_2}^2+\sigma_{\epsilon}^2}} \\
			  & = \frac{E\left[(P_1-\mu_{P_1})(P_2-\mu_{P_2})\right]}{\sigma_{P_1}\sqrt{\sigma_{P_2}^2+\sigma_{\epsilon}^2}} + \frac{E\left[(P_1-\mu_{P_1})\epsilon\right]}{\sigma_{P_1}\sqrt{\sigma_{P_2}^2+\sigma_{\epsilon}^2}}\label{eq:corr_noise_step}.
\end{align}
Since $E\left[\mu_{P_1}\epsilon\right]=0$ (the noise has zero mean) and $E\left[P_1\epsilon\right]=0$ if we consider that the noise is uncorrelated with $P_1$, the second term on the right-hand side of Equation~\ref{eq:corr_noise_step} is zero. The Pearson correlation is then given by
\begin{align}
\rho_{X_1X_2} & = \rho_{P_1P_2}\frac{\sigma_{P_2}}{\sqrt{\sigma_{P_2}^2+\sigma_{\epsilon}^2}} \\
			  & = \frac{\sigma_{P_2}}{\sqrt{\sigma_{P_2}^2+\sigma_{\epsilon}^2}} \\
			  & = \frac{1}{\sqrt{ 1+\frac{\sigma_{\epsilon}^2}{\sigma_{P_2}^2} }}.\label{eq:pearson_measured}              
\end{align}

\begin{figure}[htb]
  \begin{center}
  \includegraphics[width=\linewidth]{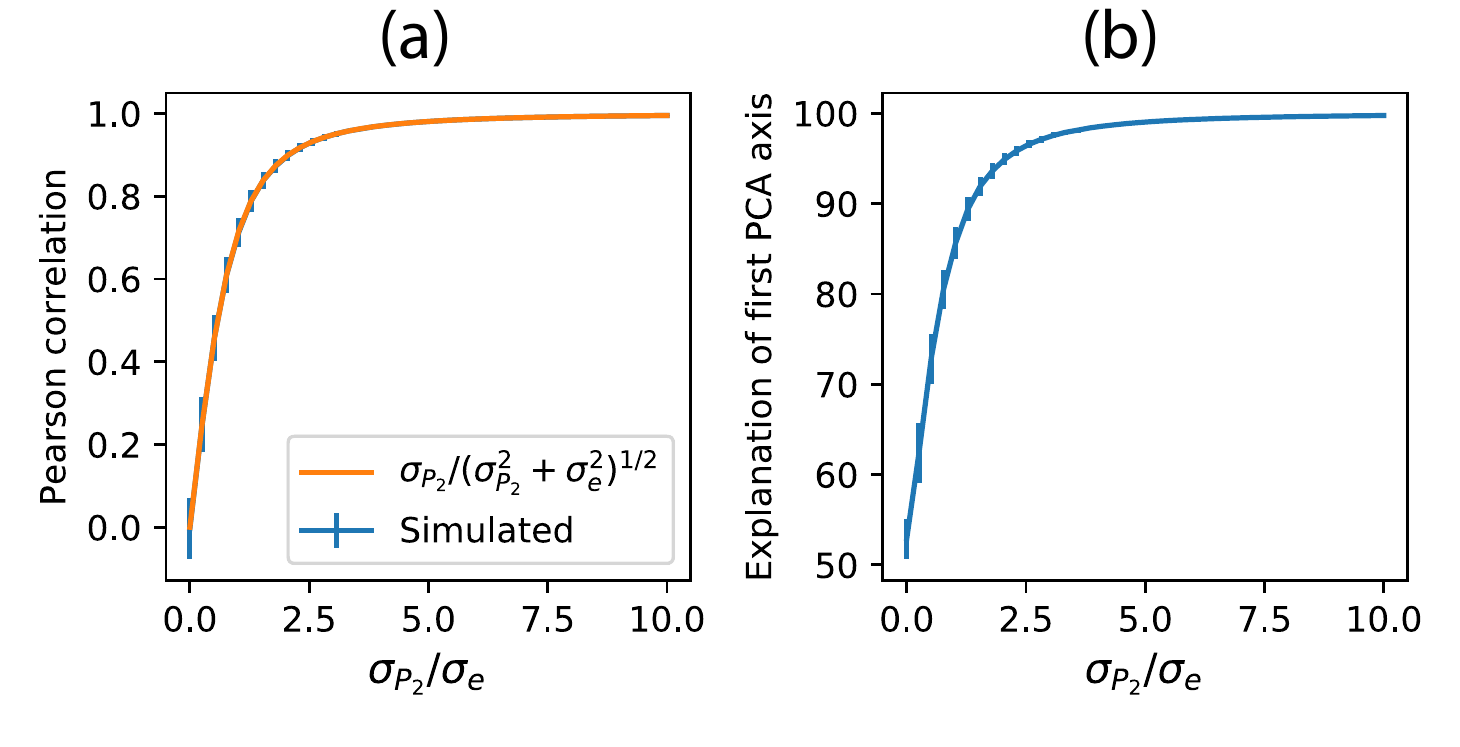} \\
  \caption{The influence of noise on data standardization prior to PCA. (a) Pearson correlation coefficient between two variables, one of them containing additive noise. The plot shows both analytical values (orange) and simulated ones (blue). (b) Explanation of the first PCA axis after applying it to the standardized variables. Error bars indicate the standard deviation obtained for 1000 realizations of the simulation.}
  ~\label{f:pearson_measured}
  \end{center}
\end{figure}

Therefore, the Pearson correlation coefficient between the two perfectly correlated variables $P_1$ and $P_2$ will be measured as $\rho_{X_1X_2}$, given by Equation~\ref{eq:pearson_measured}. Note that $\rho_{X_1X_2}$ only depends on the ratio of  the variances. Figure~\ref{f:pearson_measured}(a) shows a plot of Equation~\ref{eq:pearson_measured}, together with simulated data containing 200 objects having perfectly correlated properties $P_1$ and $P_2$, but with measured properties $X_1=P_1$ and $X_2=P_2+\epsilon$. The standard deviation of the noise was set to $\sigma_{\epsilon}=0.5$. The figure shows that when $\sigma_{P_2}/\sigma_{\epsilon} \gtrapprox 2$, the measured Pearson correlation is close to the true value of 1. As $\sigma_{P_2}/\sigma_{\epsilon}$ decreases, or equivalently, as the noise dominates the variation observed for the measurement, the Pearson correlation goes to 0.

Figure~\ref{f:pearson_measured}(b) shows the explanation of the first PCA axis as a function of $\sigma_{P_2}/\sigma_{\epsilon}$. The variables were standardized before the application of PCA. The result shows an important aspect of variable standardization: if the typical variation of the measurement is moderately larger than any variations caused by noise, the respective variable can be standardized. Otherwise, standardizing the variable may be detrimental to PCA. For extreme cases, when noise completely dominates the measurement, the obtained PCA values will indicate that this meaningless measurement has a great importance for the objects characterization.

\section{Other Aspects of PCA}

There are some important issues that need to be borne in mind when applying PCA.  These include underdetermination of the direction of the PCA axes, the stability of the transformation matrix, and the interpretation of the relative importance of the original variables.  These issues are discussed as follows.

\subsection{PCA axes direction}

\begin{figure}[]
  \begin{center}
  \includegraphics[width=\linewidth]{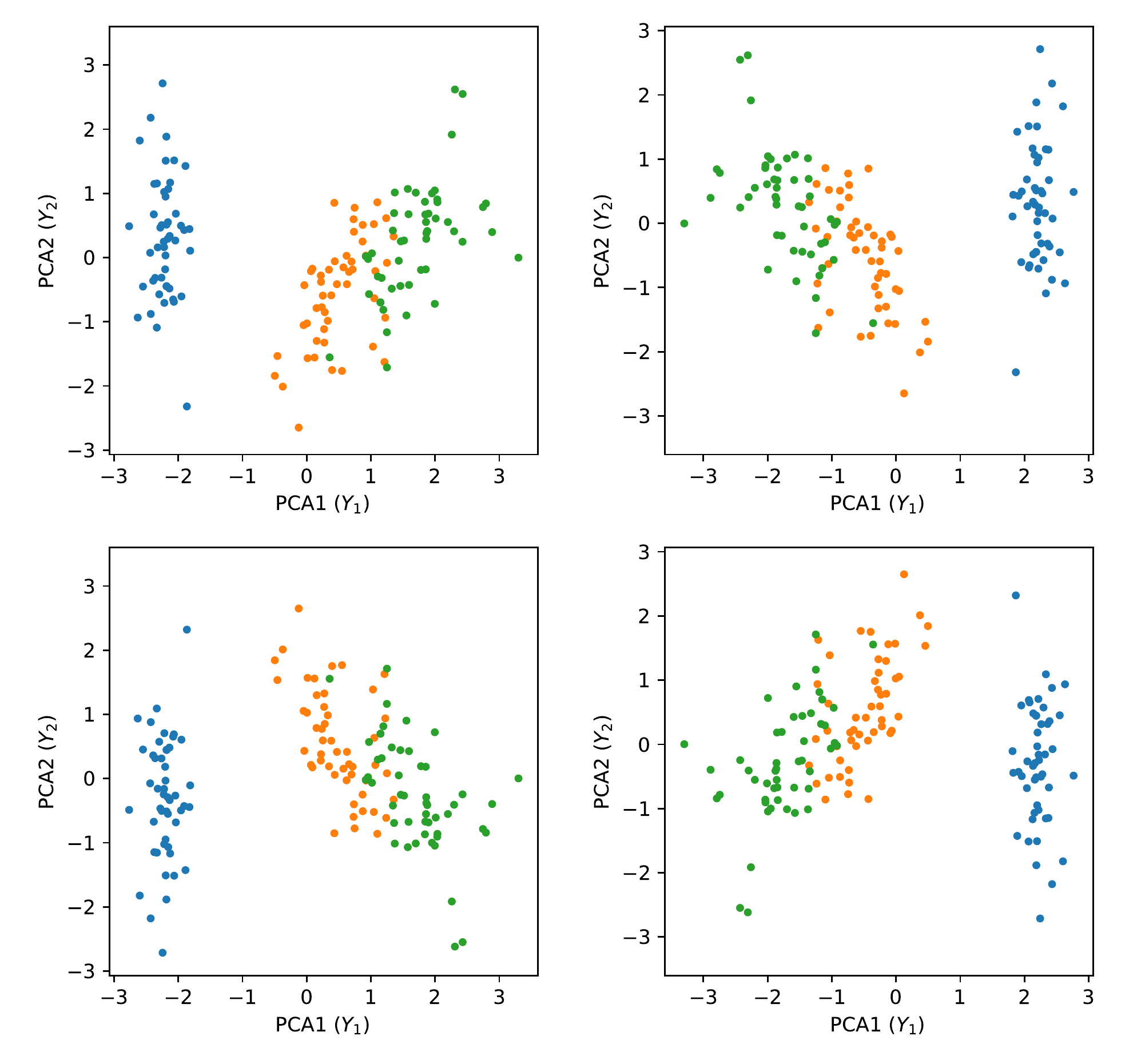} \\
  \caption{Four two-dimensional PCA projections can be obtained for the Iris dataset or, indeed, any other data. The colors identify the three categories in the Iris dataset and are included here only for reference, being immaterial to PCA.}
  ~\label{f:iris_direction}
  \end{center}
\end{figure}

It may come as a surprise to know that the directions of any of the PCA axes are not determined.  This follows immediately from the fact that if $\vec{v}$ is an eigenvector of a matrix A, so is $-\vec{v}$.  This property implies that any of the PCA axis can have its direction changed without incurring in any error.  In other words, the PCA axes directions become \emph{arbitrarily} defined.  Figure~\ref{f:iris_direction} illustrates this interesting and important property of PCA.  This figure shows the four possible PCA projections of the Iris dataset~\cite{fisher1936use} into two dimensions.  The same is true for any dataset.  Remarkably, any of the PCA diagrams in this figure are correct and are, indeed, alternative one another.

\subsection{PCA and Rotation}
\label{s:decorrDemonst}

A rotation of the data matrix $X$ is a linear transformation:
\begin{equation}\label{eq:grot}
Y=WX,
\end{equation}
where the rotation matrix $W$ is an orthogonal matrix. The covariance matrix of $Y$ is given by:
\begin{align}
Cov(Y)	&=\frac{1}{Q-1} (Y-\vec{\mu}_Y\vec{h}^T)(Y-\vec{\mu}_Y\vec{h}^T)^T \nonumber\\ 
        &=\frac{1}{Q-1} (WX-W\vec{\mu}_X \vec{h}^T)(WX-W\vec{\mu}_X \vec{h}^T)^T \nonumber\\
        &=\frac{1}{Q-1} W(X-\vec{\mu}_X \vec{h}^T)(X-\vec{\mu}_X \vec{h}^T)^T W^T \nonumber\\
        &=W Cov(X) W^T \nonumber\\
        &=W Cov(X) W^{-1} \label{eq:covrel},
\end{align}
where $\vec{h}^T$ is a column vector filled with ones and the identities $\vec{\mu}_Y=W\vec{\mu}_X$ and $WW^T=I$ were used.
The eigenvalues and eigenvectors of matrix $cov(Y)$ are given by:
\begin{align}
Cov(Y)\vec{v}_y=\lambda \vec{v}_y,\\
W Cov(X) W^{-1} \vec{v}_y=\lambda \vec{v}_y,\\
Cov(X) W^{-1} \vec{v}_y =\lambda_i W^{-1} \vec{v}_y, \\
Cov(X) \vec{v}_x = \lambda \vec{v}_x,
\end{align}
where $\vec{v}_y$ and $\vec{v}_x$ are eigenvectors of, respectively, matrices $Y$ and $X$. So the eigenvalues of the covariance matrix are conserved under rotation.

In the special case when $W$ is the eigenvector matrix of $Cov(X)$, that is, each row of $W$ contains a respective eigenvector of $Cov(X)$, $Cov(Y)$ is a diagonal matrix. If the eigenvectors are sorted according to the respective eigenvalues in decreasing order, $W$ is the PCA transformation matrix $W$. 


\subsection{Demonstration of Maximum Variance}
\label{s:maxVar}

In Section~\ref{s:decorrDemonst} it was shown that PCA decorrelates the data. It can also be shown that the PCA transform maximizes the data variance on the resulting axes. Consider a matrix W given by

\begin{equation}
Y=WX.
\end{equation}
Each row of $Y$ corresponds to a new feature after the transformation, so in order to maximize the variance of this row we need to maximize the variance of 
\begin{equation}
\vec{Y}_i=\vec{W}_i X,
\end{equation}
where $\vec{Y}_i$ and $\vec{W}_i$ are, respectively, the $i$th rows of $Y$ and $W$. From (\ref{eq:covrel}) we have:
\begin{equation}
Var(Y_i)=\vec{W}_i Cov(X)\vec{W}_i^T.
\end{equation}

In order to maximize $\vec{W}_i Cov(X)\vec{W}_i^T$ we need to constrain $\vec{W}_i$, otherwise we obtain the trivial solution where $\vec{W}_i$ is infinite. Here, we set the constraint that $W$ consists of a rigid rotation transformation, and thus it is an orthonormal matrix. Recall that a matrix is orthonormal if, and only if, its lines form an orthonormal set (i.e., $WW^T = I$). So $\vec{W}_i\vec{W}_i^T$ must be unitary.

To maximize $\vec{W}_i Cov(X)\vec{W}_i^T$ subject to $\vec{W}_i\vec{W}_i^T$ we use the Lagrange multipliers technique, maximizing the function
\begin{equation}
f(\vec{W}_i,\lambda_i)=\vec{W}_i Cov(X)\vec{W}_i^T-\lambda_i(\vec{W}_i\vec{W}_i^T-1),
\end{equation}
with respect to $\vec{W}_i$. Differentiating and equating to zero yields:
\begin{align}
\frac{df(\vec{W}_i,\lambda_i)}{d\vec{W}_i}=[Cov(X)+Cov^T(X)]\vec{W}_i^T-2\lambda_i\vec{W}_i^T=0,\\
2Cov(X)\vec{W}_i^T-2\lambda_i\vec{W}_i^T=0,\\
Cov(X)\vec{W}_i^T=\lambda_i\vec{W}_i^T.
\end{align}

We see that $\vec{W}_i^T$ is an eigenvector of $Cov(X)$, thus $W$ is formed by combining the eigenvectors of $Cov(X)$ row-wise. The respective variances are calculated as:
\begin{equation}
\vec{W}_iCov(X)\vec{W}_i^T=\vec{W}_i\lambda_i\vec{W}_i^T=\lambda_i.
\end{equation}
So, the eigenvector corresponding to the largest eigenvalue of $Cov(X)$ is placed in the first row of $W$, the eigenvector associated to the second largest eigenvalue is place on the second row, and so on. As a result, the first $M$ eigenvectors will lead to an $M$-dimensional space that posses optimal preservation of the variance in the original data.

It is interesting to observe that the efficacy of PCA in explaining variance is, to a good extend, a consequence of two properties of the eigenvectors associated to each principal axis.  First, we have that these eigenvectors are orthogonal (as a consequence of the covariance matrix being symmetric).  Then, we also have that each eigenvector corresponds to a `prototype' of the data, in the sense of having a significant similarity with the original data. As a consequence, if a given data has large scalar product with one of the eigenvectors (i.e. the data aligns with one of the eigenvectors), it will necessarily be different from the other eigenvectors as a consequence of the latter being orthogonal.  This means a substantial decay of variance along the subsequence principal axes.

\section{PCA Loadings and Biplots}

The principal axes identified by PCA are linear combinations of the original measurements.  As such, an interesting question arises regarding the identification of how those measurements are related to the implemented projection.  For instance, in the case of the beans example in Section~\ref{s:introduction}, we have that the first principal variable is defined as $PCA1 = 0.82 Diameter + 0.57 \sqrt{Area}$.  In other words, we have that the \emph{weights} of the $Diameter$ and $\sqrt{Area}$ measurements are 0.82 and 0.57, respectively.  As a consequence, the two original measurements contribute almost equally to the first principal axis.

\begin{figure*}[]
  \begin{center}
  \includegraphics[width=0.8\linewidth]{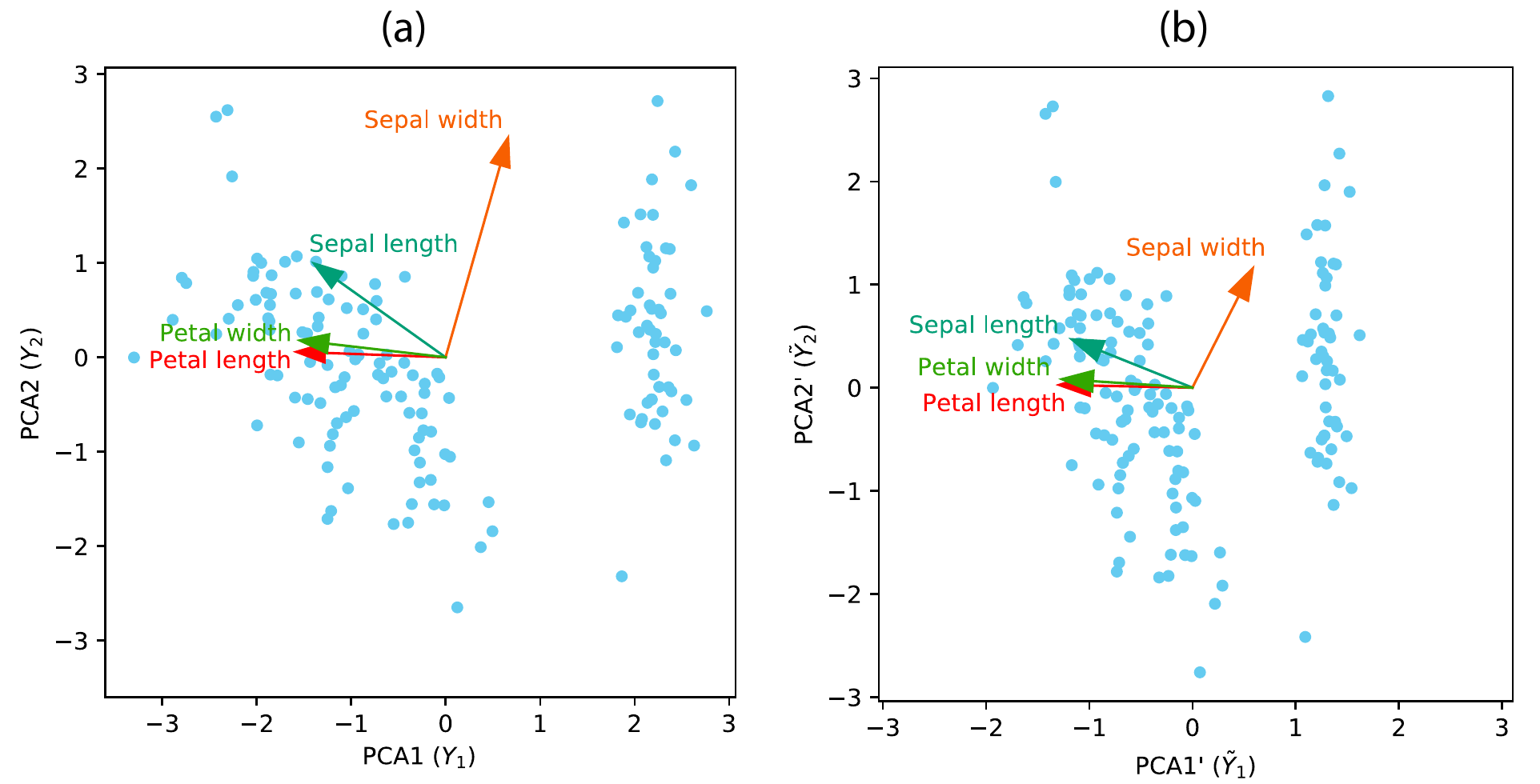} 
  \caption{Visualizing the original variables of the Iris dataset on the respective PCA projection. (a) Vectors representing the projections of the original variables onto the PCA components. (b) Biplot containing normalized PCA components and measurements vectors.}
  ~\label{f:biplot_iris}
  \end{center}
\end{figure*}

A possible manner to visualize the relationship between the original and principal variables consists in projecting the former into the obtained PCA space.  Figure~\ref{f:biplot_iris}(a) illustrates such a projection with respect to the Iris database~\cite{fisher1936use}. This database involves four measurements for each individual, namely: sepal length, sepal width, petal length and petal width.  The projections of the axes defined by each of these four original variables are identified by the four respective vectors in the PCA space in Figure~\ref{f:biplot_iris}(a).  Each of these projected vectors are obtained by multiplying the PCA matrix (eigenvectors) by the respective versor associated with the measurement. For instance, in the case of the sepal length variable, the projected vector is calculated as:
\begin{align}
\begin{bmatrix}
    \textrm{Sepal length}_1 \\
    \textrm{Sepal length}_2 
  	\end{bmatrix}
    =
  	\begin{bmatrix}
    W_{11} & W_{12} & W_{13} & W_{14} \\
    W_{21} & W_{22} & W_{23} & W_{24} 
  	\end{bmatrix}
  	\begin{bmatrix}
    1 \\
    0 \\
    0 \\
    0 
  	\end{bmatrix}    
\end{align}

Two interesting relationships can be inferred from Figure~\ref{f:biplot_iris}(a). First, we have that the angles between the projected measurements indicate relationships between the original measurements.  For instance, the fact that the petal length and petal width axes resulted almost parallel indicates that these two measurements are very similar one another.  The second relationship involving the projected variables regards their comparison with the new variables $PCA1$ (horizontal axis) and $PCA2$ (vertical axis).  For instance, we have from Figure~\ref{f:biplot_iris}(a) that the petal length is inversely aligned with $PCA1$, while the sepal width is almost parallel to the vertical axis ($PCA2$).

A closely related manner to study the relationship between original and new variables is based on the concept of \emph{biplot}~\cite{Gabriel1971biplot}.  Figure~\ref{f:biplot_iris}(b) illustrate the biplot obtained for the iris dataset. There are two main differences between the biplot and the projection shown in Figure~\ref{f:biplot_iris}(a). First, we have that the axes of the biplot are the PCA components divided by the respective standard deviation, i.e.
\begin{eqnarray}
\tilde{Y_1}=\frac{Y_1}{\sigma_{Y_1}} \\
\tilde{Y_2}=\frac{Y_2}{\sigma_{Y_2}}
\end{eqnarray}

The other difference is that the projections of the original variables are obtained by multiplying a normalized version of the PCA matrix by the respective versors, that is
\begin{align}
&
\begin{bmatrix}
    \textrm{Sepal length}_1 \\
    \textrm{Sepal length}_2 
  	\end{bmatrix}
    =   \\
  	& \begin{bmatrix}
    \sqrt{\lambda}_1 W_{11} & \sqrt{\lambda}_1 W_{12} & \sqrt{\lambda}_1 W_{13} & \sqrt{\lambda}_1 W_{14} \\
    \sqrt{\lambda}_2 W_{21} & \sqrt{\lambda}_2 W_{22} & \sqrt{\lambda}_2 W_{23} & \sqrt{\lambda}_2 W_{24} 
  	\end{bmatrix}
  	\begin{bmatrix}
    1 \\
    0 \\
    0 \\
    0 
  	\end{bmatrix}    
\end{align}
The motivation for this normalization comes from the fact that the Pearson correlation between PCA component $Y_i$ and variable $\tilde{X}_j$ is given by
\begin{equation}
PCorr(Y_i, \tilde{X}_j) = \sqrt{\lambda_i} W_{ij}
\end{equation}
Please refer to Appendix~\ref{s:biplot_calc} for a demonstration of this property. Therefore, the projection of each vector shown in Figure~\ref{f:biplot_iris}(b) onto a PCA axis correspond to the Pearson correlation coefficient between the respective measurement and the PCA component. The vector $\vec{L}_i=(\sqrt{\lambda}_i W_{i1}, \dots, \sqrt{\lambda}_i W_{i4})$ is called the loading of the $i$th PCA component. Furthermore, the angles between the vectors representing the measurements approximate well the correlations between them~\cite{Gabriel1971biplot}. Thus, the biplot provides an intuitive visualization of the relationships among the original measurements and between those and the PCA components.

\section{LDA -- Another Projection Method}


Linear Discriminant Analysis~\cite{duda2012pattern, da2009shape} - LDA is a statistical projection closely related to PCA.  It is used over categorized data, i.e.~ each of the original objects or individuals have specific assigned \emph{categories} or \emph{classes}.  As such, LDA is a \emph{supervised method}, whereas PCA is said to be \emph{unsupervised}~\cite{duda2012pattern}.  The objective of LDA is to maximize the separation between the original groups according to scatter distances defined from scatter matrices that are analogous to the covariance matrix~\cite{da2009shape}.  Because of its relative simplicity, the LDA method is completely presented in this section.

Let the original dataset contain $C$ different groups or categories. The scatter matrix for the group $C_j$ is defined as:
\begin{equation}
S_j=\sum_{i \in C_j}(\vec{\mathscr{X}_i}-\vec{\mu}_{C_j})(\vec{\mathscr{X}_i}-\vec{\mu}_{C_j})^T,
\end{equation}
where $\vec{\mathscr{X}_i}$ contains the measures for object $i$, $C_j$ is the set of objects in the $j$th category and $\vec{\mu_{C_j}}$ is the average vector for the category. The intra-group scatter matrix, $S_{intra}$, measures the combined dispersion in each group and is defined as:
\begin{equation}
S_{intra}=\sum_{j=1}^{K} S_j,
\end{equation}
where $K$ is the number of groups. The inter-group scatter matrix, $S_{inter}$, measures the dispersion of the groups (based in their centroids) and is defined as:
\begin{equation}
S_{inter}=\sum_{j=1}^{K}Q_j(\vec{\mu_{C_j}}-\vec{\mu}_X)(\vec{\mu_{C_j}}-\vec{\mu}_X)^T,
\end{equation}
where $Q_j$ is the number of objects belonging to the $j$th group and $\mu_X$ the average vector of the data matrix $X$. The matrix $S$ can now be defined as the product of the inter-group scatter matrix by the inverse of the intra-group scatter matrix, i.e:
\begin{equation}
S=S_{intra}^{-1}S_{inter}.
\end{equation}

A measurement of the separation of the groups can be readily obtained from the trace of matrix $S$ (other approaches can be used to derive alternative separation distances~\cite{fukunaga1990introduction}).  

LDA consists of applying the same sequence of operations as PCA, but with the matrix $S$ being used in place of the covariance matrix.

\section{Review of PCA Applications}

\subsection{Biology}
Data in biology come in various forms, from measurements of jaw length in vertebrates~\cite{fish2011satb2} to gene expression patterns in cells~\cite{birnbaum2003gene}. In many of these cases, the original feature space is high-dimensional, with as many as $\sim 20 000$ dimensions in the case of a gene expression profiling~\cite{Dai2005microarray,Alberts2014}. It is no surprise that dimensionality reduction methods, PCA among them, can be frequently found in many areas of quantitative biology.

In bioinformatics, high-throughput measurements of gene expression, DNA methylation, and protein profiling are on the rise as methods of exploring the complex mechanisms of cellular systems. Given a large number of measurements, in any such experiment, dimensionality reduction algorithms rapidly became an intrinsic part of the exploratory analysis in the field. As concrete examples, one may cite the \emph{arrayQualityMetrics} software~\cite{KaufmannAQC}, used for quality assessment of microarray gene expression profiling, or usage of the biplot variation for determining which variables contribute most to the samples' variance~\cite{Chapman2002biplot}. At other times, a researcher is interested in removing redundancy from his dataset before analyzing it, and thus performs a PCA before feeding his data to some more ellaborated algorithm~\cite{Lin2016confound,Wagner2015gopca}.

One might recall that a biplot refers to the practice of showing the original data axes as projected onto the principal components~\cite{Gabriel1971biplot,Gower2011biplot}. In this application, the original axes correspond to expression values of particular genes; thus, axes projected closer to the PCs indicate that the corresponding gene is strongly represented in that principal component. It serves as a clue as to which genes influence the divergence between samples.

In a considerably separate area of bioinformatics, namely structural biology, the principal component analysis technique is used to identify large-scale motions in a biomolecule's dynamics, such as protein and RNA folding~\cite{Amadei1993protein}. After calculating atomic displacements between different conformers (i.e., locally stable structures) of the molecule, covariances between motions are calculated, and the principal components provide insights into the significant structural changes~\cite{Perard2013structure}. Han et al.~\cite{Han2017phospho} applied PCA, among other methods, to understand how phosphorylation (the addition of a phosphate group to an amino acid, one of the most common post-translational modifications in proteins) induces conformational changes in protein structure.

In quantitative genetics, correlations between phenotypical features are of central importance -- e.g., the correlations between lengths and widths of certain bones~\cite{Chase2002skeleton}. The Breeder's Equation tells us that a set of phenotypic traits respond jointly to selective pressure according to the correlations between them~\cite{Lande1983,Solomon2006QTL,Steppan2002Gmatrix}. Thus, PCA serves as a way to extract the directions along which significant evolutionary changes are more likely to happen and visualize them directly.

Population genetics, on the other hand, deals with prevalences of certain genotypes in a population, or preserved sequences between groups. Here, PCA is applied to genetic variation data in various groups and species, and used to identify population structures~\cite{Byun2017ancestry} and putative migratory or evolutionary events~\cite{Reich2008genetics,novembre2008genes,hofmanova2016early,patterson2006population}. An interesting application by \emph{Galinsky~et~al.}~\cite{Galinsky2016alcohol} analyzed the allele distribution of a population and compared its principal components to those of a null distribution derived from a neutral model, identifying genes undergoing natural selection in a population. In particular, they observed that ADH1B, an enzyme associated with alcohol consumption behaviors, seems to be undergoing simultaneous and independent evolution in both Eastern Asia and Europe.

In ecology, one might be interested in comparing data from several different species~\cite{Ramette2007micro,Giannini2011niche}. For instance, microbial ecology is often concerned with metabolic profiles or gene expression of various microorganisms in the same environment, leading to a dataset where variables are concentrations of a specific catabolite and samples indicate different species in a substrate. In another scale, PCA of transect data (i.e., counting the occurrence of certain species along a predefined path) is a conventional approach to distinguish between different animal communities~\cite{Huettmann2001transect}. In another example, PCA was employed to reduce the amount of data in a Maximum Entropy model to infer the distribution of red spiny lobster populations in the Galapagos Islands~\cite{Moya2017lobster}.

\subsection{Medicine}
Modern medical science relies on sophisticated imaging techniques such as functional Magnetic Resonance Imaging (fMRI), Positron Emission Tomography (PET) and Computed Tomography Imaging (CTI)~\cite{Nandi2015medical}. The obtained data need to be pre-processed: noise must be extracted, redundancies must be discarded, and different sources are to be gathered~\cite{Priya2015denoising}. Thus, PCA is often used in these steps as a computationally efficient and yet reliable technique to aid in producing the image.

Apart from imaging, several clinical variables may be combined to maximize the information obtained from a patient's data like age~\cite{Jolliffe1992medres}, concentrations of certain substances in the blood~\cite{Agarwal2012metabolic}, Glasgow scores~\cite{Koziol1990glasgow}, electrocardiogram (ECG)~\cite{Martis2012ECG} signals and others. These may then be used to classify the patient's possible outcome, and in this process, it may be necessary to rotate or combine variable axes~\cite{Jolliffe1992medres,Martis2012ECG,polat2007expert,polat2007detection}. 

Another type of application that has incorporated PCA is related to diagnostics.  For instance, PCA was used to reduce the dimensionality of the data in the diagnostic prediction of cancers~\cite{khan2001classification}. In this way, gene-expression signatures were analyzed, and artificial neural networks were employed as a classifier. Chemical-related tools were also employed with PCA in the diagnosis approaches, such as in~\cite{yang2004diagnosis}, in which the authors proposed a methodology to improve the differentiation between hepatitis and hepatocirrhosis. For that, metabolites that take part of samples of urine were analyzed.

\subsection{Neuroscience}

The brain is a structure with an extremely high number of components and a very complex topology. Therefore, statistical tools, including the PCA, are useful for studying it. In neuroscience, PCA is often used in classification methods and data analysis of measurements of brain activity and morphology, such as in electroencephalography (EEG) and Magnetic resonance imaging (MRI). PCA is also used as a data analysis tool of psychophysical experiments.

Epilepsy is a neurological disorder that is associated with uncontrolled neuronal activity which may lead to seizures. The electroencephalography (EEG) is often used for epilepsy diagnosis and seizures detection through identification of EEG markers or abnormalities. Epilepsy diagnosis is a complicated task, so the diagnosis is normally confirmed by the EEG interpretation by a neurologist while taking into account the medical history of the patient. Due to possible error in the diagnosis, it is interesting to have a precise automatic system that could assist epilepsy diagnosis and the detection of seizures. 

In \cite{ghosh2008principal} the authors propose a supervised classification method that consists of PCA applied to nine selected features of the EEG. The transformed data serves as input of a cosine radial basis function neural network (RBFNN) classificator. The method could classify the patient EEG in normal, interictal (period between seizures) and ictal (during a seizure), with a false alarm seizure detection of 3.2\% and a missed detection rate of 5.2\% for the parameters and data utilized in the article.

Magnetic resonance imaging (MRI) is an imaging technique capable of obtaining high quality pictures of the anatomy and physiological processes of the human body, including the brain.  MRI is widely used for clinical diagnosis. In the case of some neurological diseases, the diagnosis is sometimes asisted by an automated classification based on the brain MRI image. In \cite{zhang2012mr}, a classification method is proposed for MRI images that employs PCA after the discrete wavelet transform (DWT). The PCA reduces the dimension of the feature space from 65536 to 1024 with a 95.4\% of the variance. The PCA processed data is used as an input of a kernel support vector machine (KSVM) with the GRB kernel, so as to infer the health of the brain. The diseases considered in the method are the following: glioma, meningioma, Alzheimer disease, Alzheimer disease plus visual agnosia, pick disease, sarcoma, and Huntington disease.

The dendrites of a neuron can grow in a very complex and branched way. The respective arborizations can be digitalized as a set of points in a 3D space representing its roots, nodes, tips and curvatures.  PCA can be used to describe a dendritic arborization \cite{yelnik1983principal}. The shape of the arborization can be described by the relative values of the standard deviations of the new digitalized dendritic arborization data after PCA application. The dimensions of the arborization are defined as the length of the interval between the most extreme points projected in each of PCA axis. Depending on the shape of the dendritic arborization, the PCA axes are utilized to determine the orientation of the arborization.

The accuracy of the ability to recall past painful experiences is still object of controversy. A generally accepted way of describing pain is by a sensory-discriminative (intensity) and affective-motivational (unpleasantness) dimensions. In \cite{khoshnejad2014remembering}, the authors conduct a psychophysical experiment to study the ability to recall pain intensity and unpleasantness in a very short time interval. The subjects were thermally stimulated and evaluated in real time. The intensity and unpleasantness of the pain was inferred by a visual analog scale (VAS) both simultaneously and in a short time after the stimulation. The PCA is used over the VAS data and the first three principal components are used for further analysis, explaining about 90\% of the variance. The results of the study support the loss of pain memory information and reveals a significant difference in the ability to recall the stimuli between the subjects.


\subsection{Psychology}
In psychology-related areas, quantitative data is often provided in the form of examination scores. Part of its information can be analyzed in terms of multidimensional statistics. For instance, PCA was employed in the analysis regarding how people store information through a memory test~\cite{Paulino2016memory}. The researchers considered the childhood and adolescent development and found differences related to age and the cognitive maturation process.

Efforts directed to better understanding of behavior can also employ PCA. For instance, studies regarding the connection between memory and anxiety~\cite{beuzen1995link}, and the relationship between the organization of working memory and cognitive abilities~\cite{alloway2004structural}. Facial expressions were also investigated by using a PCA-based approach~\cite{calder2001principal}. 
This study considered datasets of faces and obtained features from the considered images. The results indicate that pictures of facial expression, when processed by the PCA-based approach, provided reliable results when compared to the social psychologist's analysis.

\subsection{Sports}
The existence of multivariate measurements in sports provide many opportunities for PCA applications. For instance, the precision required by elite athletes and martial artists requires coordination between several parts of the body, and kinematics-derived measurements can be submitted to a PCA in order to undercover synergies and principles of a specific sportive practice~\cite{Zago2017karate,Gloersen2018technique}, or pinpoint health-hazardous practices in everyday actions such as walking~\cite{Baudet2014gait}. Furthermore, PCA can be employed in tests of dopping~\cite{sottas2006statistical,norli1995chemometric}, such as in~\cite{norli1995chemometric} in which the authors employed PCA as a dimensionality reduction to the measures of anabolic steroids. 

PCA has also been applied to compacting three-dimensional coordinates of body points at different times~\cite{federolf2014application}.  In~\cite{federolf2014application} PCA was applied to 26 three-dimensional body coordinates of 6 alpine ski racers. In this experiment, the first four principal components were responsible for 95.5\% of the PCA variance.  In order to study the performance of vertical jumping, researchers considered PCA to eliminate correlation in the athlete's data measurements. Another analyzed characteristics of athletes that employed PCA are related to somatic anxiety, which means the physical symptoms of anxiety~\cite{smith1990measurement}.

\subsection{Chemistry}
\label{sec:chemistry}
Some analyses in chemistry-related areas have to deal with a large amount of data. For instance, in analytic chemistry, data can be generated from the analysis of samples through different types of equipment, such as NMR (Nuclear Magnetic Resonance)~\cite{de2017mate,nord2001multivariate}, EPR (Electron Paramagnetic Resonance)~\cite{serudo2007reduction}, Mass Spectrometry~\cite{kelly2005tracing,serra2005determination}, etc. These experiments normally generate signals as output, which are then analysed in order to search for patterns.

A possible manner to find patterns and structure in these sets of data is by visual inspection, in which the skills of the operator can strongly influence the analysis. In order to achieve a more controlled and comprehensive analysis of the measured data, concepts of multivariate statistics and data analysis have been incorporated, giving rise, around the 70s, to a new area called \emph{chemometrics}~\cite{varmuza2016introduction,geladi1990start,tauler2009comprehensive}. The consolidation of Chemometrics as a research area was promoted in 1974, when \emph{Bruce Kowalski} and \emph{Svante Wold} started the Foundation of Chemometrics Society~\cite{geladi1990start}.

These techniques are normally used in studies related to data-driven methods, in which empirical methods are employed~\cite{varmuza2016introduction}.  One of the most important tools of chemometrics is the PCA technique~\cite{bro2014principal}, which has been incorporated into many studies, including the analysis of food~\cite{forveffle1996multivariate,de2017mate}, drugs~\cite{bailey2002multi,wang2004metabolomic}, disease diagnosis~\cite{ouyang2011metabolomic}, the presence of pollutants in water~\cite{serudo2007reduction}, etc. The use of PCA in chemistry-related studies is normally related to the following two main aspects: (i) data visualisation and (ii) dimensionality reduction.

A possible application of PCA in food chemistry regards the data visualisation of metabolomic analyses~\cite{de2017mate,ceribeli2018mate}. In a recent study, the quality of cattle meat was characterised with respect to different diets~\cite{de2017mate}. The animals were grouped into different classes fed with different amounts of mate herb extract. Levels of metabolites in the meat were measured using \textsuperscript{1}H~NMR technique and PCA was employed to better understand the relationship between meat quality and the animal feeding. As data were projected onto principal components, the classes emerged naturally. Furthermore, in order to understand the relationship among the different metabolites, the concept of loadings was used.  Other works investigating food chemistry have been reported~\cite{forveffle1996multivariate,nord2001multivariate}, including the use of PCA as an auxiliary method to classify different types of grapevines~\cite{forveffle1996multivariate}. 

Other applications in Chemistry include the diagnosis of diseases, such as identification of pancreatic cancer in patients~\cite{ouyang2011metabolomic}. More specifically, the patient serum was analyzed and PCA was employed as a data reduction method~\cite{ouyang2011metabolomic}, providing support for multivariate analysis. In other studies, the PCA technique was applied in order to identify the chemical characteristics of phytomedicines~\cite{bailey2002multi,wang2004metabolomic}. Samples prepared from river water were analyzed by EPR and the data was then projected in PCA~\cite{serudo2007reduction}. Different indications of the mercury cycle were found along the river.

\subsection{Materials Science}
Many types of equipment can be used in order to describe characteristics of samples in material science. Normally the focus is on the design and discovery of solid materials, such as ceramics~\cite{scott2007prediction}, polymers~\cite{eynde1997tof}, and others~\cite{shenai2012applications,wagner2001characterization,rajan2005materials}. These types of materials can also be described in terms of multivariate data and data analyses. 
Consequently, PCA has become an important part of such investigations~\cite{rajan2005materials}. 

In order to probe the efficiency of PCA in the materials science area, the authors of~\cite{suh2002application} studied the problem of multivariate analysis underlying such a technique. A case study about superconductors was proposed, in which the authors analyzed the data provided from this type of material by employing PCA. More specifically, conductivity characteristics were measured for samples submitted to high temperatures. In order to illustrate how meaningful the obtained PCA could be, some characteristics were explored and the authors concluded that this technique provided potential descriptors to be used in materials science. 

Similar analyses have been done in order to characterise polymers, such as the case of Polystyrene~\cite{eynde1997tof}, in which PCA was applied to the signal quantification stage. More specifically, the authors used the Time-of-Flight Secondary Ion Mass Spectrometry (ToF-SIMS) and obtained the samples spectra, with more than $150$ peaks. In order to aggregate the information of all these spectra, PCA was used with promising results.  Another application that considered spectra from the same kind of equipment, ToF-SIMS, is the characterization of adsorbed protein films~\cite{wagner2001characterization}. The protein spectra were measured and PCA was then applied. Though the spectra from different proteins were similar, by projecting these data onto the principal components it was possible to visually identify that the proteins were organized into distinct groups. Furthermore, PCA supported the identification of the peaks that varied more intensely.

PCA was employed in order to illustrate characteristics of nanomaterials, showing that some materials are separated into groups~\cite{tisch2010nanomaterials}. PCA was also used in ceramic characterisation.  For example, it was applied as a step to predict functional properties of ceramic materials, e.g., properties of non-metallic, inorganic, and polycrystalline materials~\cite{scott2007prediction}. In particular, PCA was used as a feature selection method capable of reducing the dimensionality of the original data. In the aforementioned study, the selected data was classified by using an artificial neural network.

\subsection{Engineering}
As a consequence of the generality of PCA, it is expected that it can provide a valuable auxiliary resource also in Engineering.  For instance, this technique has been used in electrical engineering~\cite{costa2016negative,costa2018pattern}, civil engineering and structural health~\cite{hua2007modeling,kerschen2004sensor}, and mechanical engineering~\cite{shuang2007bearing,malhi2004pca,kwan2001gearbox,liu2007gearbox,li2010gearbox}.

In civil engineering, an automatic method of guiding the management and maintenance of large-scale bridges was proposed~\cite{hua2007modeling}. This system is focused on long-term structural health monitoring systems and is based on some machine learning techniques, including PCA. One characteristic of this system is the vibration-based damage detection that can determine the presence, location, and severity of structural damage. These characteristics are measured from changes in modal parameters in the frequency domain. One of the main challenges is the temperature, which is related to the environmental conditions and can change the modal settings. The method consists in reducing the dimensionality by using PCA applied to the long-term temperature data measured from different sensors. This compressed data is used as the input of a support vector regression (SVR)~\cite{basak2007support}.  

Apart from employing PCA as a tool to preprocess the input data, this technique was also applied to evaluate the accuracy of sensors~\cite{kerschen2004sensor}. Because some sensors can fail, the input information should be validated, which is necessary to have accurate measurements. Furthermore, the proposed technique can detect, isolate and correct the faulty sensor. 

In mechanical engineering-related areas, PCA has been applied to the task of monitoring health quality of a given type of equipment or system. For instance, a methodology for fault diagnosis by considering multidimensional and temporal data regarding rolling bearings~\cite{shuang2007bearing}. PCA was employed in order to reduce the amount of data, and such compressed data was used in a machine learning method based on Support Vector Machine (SVM). Another study, which employed PCA as a feature selection method, also took into consideration the problem of bearings defects classification of machines~\cite{malhi2004pca}. By considering different classifiers, the features were able to improve the results in case of supervised and unsupervised classifications.

Another critical problem is the failure detection in gearboxes. Because a relationship between temperature and failures has been observed, an infrared camera was used to capture images in real time~\cite{kwan2001gearbox}. These images, which reflect the temperature, were analyzed through two image processing techniques. The first analysis consisted in defining an index based on measuring the growth of the heated region. The second one employed PCA as part of the approach to compute features and reduce the dimensionality. The analysis showed that the PCA-based approach is more robust regarding environmental changes. Other studies dealt with this problem by employing modified versions of PCA to analyze time series measured by the vibration of gearboxes~\cite{liu2007gearbox,li2010gearbox}.

In the industrial sector, several chemical compounds are produced, mainly in batch reactors. Some examples of these compounds are polymers, pharmaceuticals, and biochemicals. Monitoring these batch processes is a important and complex task which ensures the safety of the process and the high quality of the products. In \cite{nomikos1994monitoring} the authors reported a MPCA (Multiway PCA) based method to monitoring the progress of batch process. The method uses MPCA to create a reduced space of the historical dataset of past batches and compares at each time interval the current batch variables trajectory with the past batches data in the reduced space, thereby detecting possible anomalies. In statistical process control (SPC), the biggest advantage of this method is that it only uses data from past batches, thus not requiring the detailed knowledge of the system. 

Recently, in the field of analog electronics, PCA has been employed to better understand the parameter variability within and among different families of transistors.
In~\cite{costa2016negative}, a set of features were obtained experimentally for a collection of transistor devices of several families. Assuming the transistors to operate as amplifiers, several device performance features and parameters were estimated, including the harmonic distortion of the respective transfer curves for three levels of negative feedback. Three main results were obtained for the considered settings and devices: (i) the variation of parameters was relatively small within each transistor type, implying in respective clusters; (ii) moderate level of negative feedback was found not to be able to completely eliminate parameter variations; and (iii) high variance explanation was achieved by using only two axes.  The latter result was further developed giving rise to a new modeling approach~\cite{costa2018characterizing}.

A subsequent work~\cite{costa2018pattern} focused  on studying the patterns of parameter variability among devices encapsulated in the same transistor array. PCA was employed in a similar fashion as in the previous work to quantify parameter variability. The results confirmed the substantially higher uniformity among parameters of transistors belonging to the same array.









\subsection{Safety}

In the safety area, the study conducted in~\cite{AIC:AIC690421011} devised a model based on PCA to identify the correlation among sensors. Usually, the identification of errors in sensors can be performed in different ways. Some measurements can reach unusual values – providing indication of a failure – which can be easily identified by setting lower and upper limits. Nevertheless, some minor failures cannot be detected in such a straightforward way, since the measured absolute values do not take on extreme values. The use of correlation via PCA can be used to identify such minor errors in faulty sensors since, in a typical operation, measures obtained by sensors are usually correlated. Given the correlations in the normal operation, the PCA technique was also used to reconstruct measurements of faulty sensors via orthonormal decomposition.

\subsection{Computer Science}
In computer science, PCA has been employed in a diverse range of applications, varying from specific tasks, such as biometrics and data compression to more general problems, including unsupervised classification and visualization.

In the scope of computer vision, one of the most iconic applications of PCA is face recognition, known as \emph{Eigenface method}~\cite{turk1991eigenfaces,kirby1990application,sirovich1987low}. This technique is used to recognize faces in images by linearly projecting them onto a lower dimensional space. For this, the sequence of pixel intensities along each image is considered as the feature space, and PCA is employed to find relevant eigenvectors (named \emph{eigenfaces} in this context).

In the Eigenface method, classification is attained by PCA projecting both the unknown image and reference faces onto the eigenspace and calculating the respective Euclidean distances. Since the size of the covariance matrix grows quadratically with the number of pixels, its direct calculation is often unfeasible. 

However, because the rank of the covariance matrix is limited by the number of samples, Singular Value Decomposition (SVD) can be used directly over the feature space, dropping the need to compute the covariance matrix explicitly. Other methods similar to the Eigenface have been developed for other tasks also involving biometric data. This includes recognition of palm print~\cite{lu2003palmprint, connie2005automated}, iris~\cite{basit2005efficient}, gestures~\cite{gawron2011eigengestures} and behavior~\cite{fookes2010eigengaze,fookes2009gaze}. In~\cite{martinez2001pca}, the authors conclude that, for the task of face recognition, PCA can sometimes outperform LDA when the training dataset is small, while also being less sensitive to changes in the training data.

Another application of PCA in computer vision is the unsupervised classification of images and videos. Among the popular techniques for this task is GPCA (Generalized Principal Component Analysis)~\cite{vidal2005generalized}, which uses PCA to combine subspaces defined by homogeneous polynomials for data consisting of sets of images or video clips. Classification is attained by applying a clustering algorithm in the reduced space. The technique was found to be appropriate for many specific tasks, including segmentation of videos along the time, face classification with different illumination conditions and tracking of 3D objects in video clips. 

Other frequent applications of PCA in computer vision are based on the idea of applying PCA followed by a clustering algorithm over the raw data or sets of features extracted from images. In texture analysis~\cite{bharati2004image}, the set of features are usually obtained from data lying on the frequency space (such as subspaces obtained from Fourier or Wavelet transforms). In~\cite{ke2004pca}, PCA is used to combine image descriptors given by the SIFT method~\cite{lowe1999object}. This class of descriptors is obtained by finding points of interest in the images, resulting in a local set of descriptors for each image. The authors found that the results obtained by applying PCA, in comparison to using just the histograms of the descriptors directly, gives not only a more compact representation of the images but also significantly improves the matching accuracy of some tasks.  These tasks include tracking of objects across different images obtained from real-world or controlled three-dimensional transformations (such as rotation, shift, and scale). PCA was also employed for the image retrieval, where, for a provided query image, the algorithm finds a set of similar images from a large database~\cite{gong2013iterative}.

An interesting result regarding the use of PCA in image analysis is that certain classes of neural networks, trained with image datasets, seem to mimic the PCA transformation. This approach is partially confirmed by the fact that multilayered neural networks use uncorrelated linear projections of the data as internal representations~\cite{brunelli1993caricatural,bourlard1988auto}. In~\cite{oja1992principal}, the authors develop a mathematical proof connecting the approach taken by some neural networks with PCA, which is accomplished by creating a neural network that effectively reproduces the PCA transformation.

In general data analysis, PCA can be regarded as a pre-processing step to be applied to the data before using more sophisticated methods for classification or learning~\cite{malhi2004pca}. In such a context, PCA can also be understood as a feature selection or feature extraction process in the sense that it does not result in a subset of the first features but a combination of them. Many benefits can be attained by using this approach; for example, the computational cost of a method can be reduced substantially with minimal loss of accuracy by also reducing the size of the input data before applying a more complex classification algorithm~\cite{song2010feature, ekenel2004feature}. However, one should be careful when combining PCA and other classification techniques, as this kind of benefit depends on the classification technique being used and the dataset. For instance, in~\cite{janecek2008relationship} the application of PCA before Support Vector Machine (SVM) considerably reduced the classification accuracy of the analyzed datasets. 

Another notable application of PCA in computer science is data compression. In particular, PCA was used to compress image data. An example of this approach is present in~\cite{du2007hyperspectral}, in which the authors propose a modification of the JPEG2000 standard by incorporating an extra step based on PCA to improve the rate-distortion performance on hyperspectral image compression. Results show that PCA outperforms the traditional approach in which the coder is based on spectral decorrelation using wavelets. In a similar direction, the work~\cite{bailer1997neural} also employed PCA as a technique to compress data from stellar spectra. The results show that PCA attained a compression rate of $30:1$ while keeping $95\%$ of the variance. 

Aside from images, PCA was also employed to compress other types of data. In \cite{babenko2014neural}, PCA is used to compress neural codes into short codes that maintain the accuracy and retrieval rates similar or better than state of the art techniques. Neural codes are visual descriptors obtained from the top layers of a large neural network trained with image data. These can be used to retrieve data from large datasets. PCA was also found to be useful in compressing data describing human motion sequences~\cite{liu2006segment}. This kind of data incorporates three-dimensional trajectories of sets of markers that represent the motion of the human skeleton captured from human actors performing specific actions. The technique is based on compressing the positions of the markers to a lower dimensional space using PCA for each keyframe of the animation.

\subsection{Deep Learning}
Many of the proposed methods regarding neural networks have employed PCA as dimensionality reduction pre-processing step~\cite{oja1992principal,chen2002dynamic,sahoolizadeh2008new}. A new area of study, called Deep Learning, emerged and many related approaches have also incorporated PCA. Usually, as in the case of standard neural networks, PCA is employed in deep learning area as a pre-processing step, in which the data is reduced, and the first principal components can be used as features. Some of the studies that applied PCA are the classification of hyper-spectral data~\cite{chen2014deep}, face recognition~\cite{face2014Sun}, and to extract features from videos~\cite{zou2012deep}.

Because of the high dimensionality that takes part in hyper-spectral measurements, there is a possibility to use PCA to compress the input data. In~\cite{chen2014deep}, remote sensing data were measured by many different characteristics of analyzed regions, for instance, information taking into account trees, water, streets, among others. One of the employed pipelines involves PCA as the first step, in which PCA compresses the multidimensional data. Next, such compressed data is summarized according to the neighboring regions and is then flattened into a vector. Finally, this vector can be employed as input to the neural network. Note that this method and other variations of such pipeline were used to create features to describe the system. Classification tests by using these features were applied and, as a result, the authors found that the proposed features provided higher accuracy when compared to some other methods.

In deep learning, PCA is commonly used as part of other image analysis tasks, such as generating features concerning the face recognition methods, and this is also used together with deep learning~\cite{face2014Sun}. Another example is the PCA-based deep learning that considers a cascade of PCAs to classify images, called PCANet~\cite{chan2015pcanet}. This methodology was applied to many tasks, such as recognition of hand-written digits and objects, and the results were compared to other deep learning based approaches. In general, good results were obtained, and the authors suggest that PCANet is a valuable baseline for tasks involving a significant amount of images. 

In another study, different deep learning approaches were proposed to unsupervised classification of sleep stages~\cite{langkvist2012sleep}. This method considered PCA after the feature selection step, which was used in order to capture the most of the data variance. As a parameter, the authors used the first five principal components. In general, the reached accuracy found for the PCA-based method was not the best one. However, this automatic approach illustrated a way to classify sleep stages, without considering specialist knowledge. The proposed methodology can also be used in tasks of detecting  anomaly and noisy redundancy.

\subsection{Economy}
Most of the applications of PCA in Economy are based on evaluating the financial development of countries or entities using a combination of features~\cite{vyas2006constructing}. Usually, PCA is employed to combine economic indicators in order to attain a smaller set of values so that these entities can be ranked or compared. An example of this approach is present in \cite{hosseini2011dynamic}, in which countries are ranked according to the principal components obtained from sustainability features taken over time. The authors indicate that while some progress was found for economic development, the overall conditions got worse during the considered period. Other works also consider PCA to build an integrated sustainable development index, such as in \cite{abou2013integrated} and \cite{doukas2012assessing}.

Another example of using PCA to aggregate economical indices is explored in~\cite{fifield2002macroeconomic}, in which the loadings resulted from PCA of several indices were employed to determine the importance of macroeconomic indices for countries. The article also compared the combination of macroeconomic indices and shared returns of their respective stock markets. 

In other approaches, PCA was used to better understand the local economic characteristics of cities or provinces. In~\cite{drafor2017principal}, it was used to visualize data involving living conditions of households in rural and urban regions of Ghana. Results indicate very distinct characteristics between the population living in rural area and those in the urban region. By using a similar approach, in~\cite{wang2013spatial}, the evolution of economic characteristics of the Liaoning province are studied throughout time. A coordination development index was devised by using PCA.

\subsection{Scientometry}
Scientometric sciences are devoted to studying the qualitative aspects of science~\cite{tague1992introduction}. Typical analyses include the assessment of the scientific impact of journals, institutions, and scientist via metrics such as the total number of articles, citations, views or patents~\cite{sengupta1992bibliometrics}. Popular topics of interest in scientometric studies are the evolution of science~\cite{almeida2009science}, the identification of interdisciplinary~\cite{porter2009science} and co-authorship dynamics~\cite{liu2005co}. Because many aspects of science are subjective (e.g., the concept of quality), many measures have been proposed to capture different views. PCA, in this case, has been used as a visualization tool and, most importantly, as a way to make sense of scientometric data. 

Several indexes have been proposed to assess the quality of universities worldwide. However, the validity of some criteria has been questioned by academic stakeholders. In~\cite{Docampo2015}, the authors studied whether the metrics conceived by the annual academic rankings of world universities (ARWU, see Liu and Cheng 2005) privilege larger universities.
The authors argue that this is an essential debate because such an alleged privilege may cause institutions to pursue a growth devoid of quality since much importance is currently being given to size. They argued, via PCA analysis of several metrics in the ARWU ranking, that two main factors would account for the data variability. While the first principal component accounts for 54\% of the variance, the second component was found to explain 30\%. A more in-depth analysis of the variables revealed that, in fact, the size factor accounts for a significant variance. However, the excellence factor seems also to play an important role, as related metrics populate the first principal component.

An important point of interest for scientometric researchers concerns the introduction of measurements to quantify the relevance of research, journals, and papers. While most of the metrics rely on some type of citation information, there is no consensus on which would be the most important measurement. In addition, if a multi-view impact is desired, it would be important to understand which measurements are interrelated. In this context, a systematic comparison of 39 impact metrics was performed in~\cite{10.1371/journal.pone.0006022}. The considered metrics included traditional metrics based on raw citation and usage data, social network measures of scientific impact, and other hybrid metrics. The PCA analysis revealed that the first principal component identifies citation measures, discriminating them from almost all usage metrics. The second principal component accounts for the discrimination between citation and social network metrics. All in all, the clustering observed with the PCA projection revealed that the impact metrics could be interpreted according to two main dimensions: (i) the time in which the evaluation is performed (i.e., usage vs. citation), and (ii) the dimension discriminating popularity from prestige (citation vs. social impact).   

The detection of \emph{evergreens} through principal component analysis was performed in~\cite{ZHANG2017629}. Differently from conventional scientific papers, \emph{evergreens} are those manuscripts in which the number of citations regularly increases, with no significant decay effect over time. Even though the predictability of the consistency of evergreens is unfeasible, it is still important to understand the behavior of their citation trajectories. The method proposed in~\cite{ZHANG2017629} for clustering the citation behavior of evergreens consists in decomposing the trajectory curves via functional principal component analysis. Such a decomposition is then used for data partitioning via K-means~\cite{berkhin2006survey}. The main results suggested that most of the data variability could be explained solely by two functional components. The main component, which explains 95\% of data variation, is characterized by a steadily growing citation curve. In fact, it is related to the behavior of most evergreens. The main findings obtained by this method based on functional PCA suggest that papers with similar citation patterns shortly after their publications may display distinct trajectories in the long run.

\subsection{Physics}

Most of the current problems in physics can be approached in two main manners:  by developing the basic laws for the problem (\textit{Ab initio}), or by constructing an empirical model regarding the relationships of some aspects of the considered system, which should be confirmed at least for an experiment. Note that the second type of approach is inspired by a sequence of tests in which the parameters of the proposed model are systematically varied. When the dimension of the associated data is considerable, PCA can be applied to determine which of them are potentially more relevant~\cite{van2017learning, boon2001systematic}. 

The area of quantum many-body problems involves the study regarding many interacting particles in a microscopic system.  The solution of this problem is related to the high dimensionality of the underlying Hilbert space, which makes PCA a useful tool for distinguishing and organizing configurations in the underlying space. PCA can be applied to recognize the phases of a given quantum system.  For instance, a  neural-network-based approach has been used to identify the phase transition critical points after a preliminary PCA-assisted identification of phases~\cite{van2017learning}. PCA was also used to reveal that the energies of crystal structures in binary alloys are strongly correlated between different chemical systems~\cite{curtarolo2003predicting}. This study proposed an approach that uses this information to accelerate predicting the crystal structure of others materials.

In the same context, other studies considered PCA as a tool to better analyze quantum mechanics. In order to find insights about quantum statistical properties, a quantum correlation matrix was defined, and the  PCA computed~\cite{mosetti2016principal}. In this study, the results were discussed regarding the quantum mechanical framework. In \cite{lloyd2014quantum}, the authors introduced a quantum version of PCA, called qPCA. This method consists in a quantum algorithm that computes the eigenvectors and eigenvalues of a density matrix, $\rho$, which describes a mixed quantum system. In the case of $\rho$ being a covariance matrix, the algorithm can perform a classical PCA.

Among other areas, in nuclear physics, PCA has been applied to study the effect of many nuclear parameters on the classification of even-even nuclear structures~\cite{al2015principal}. Additionally, PCA was used to look into the local viscoelastic properties and microstructure of a fluid through a series of images that perform the motion in this fluid~\cite{chen2015principal}. More specifically, the authors considered a series of images of suspended particles in a Newtonian fluid (Brownian motion). PCA has also been used to solve problems in network theory, such as for understanding network measurements~\cite{silva2016concentric}, analyzing gene networks~\cite{couto2017effects} and analyzing and visualizing data obtained from text networks~\cite{ferraz2017representation}.

Similarly to the methods already discussed in section~\ref{sec:chemistry}, in physical chemistry, PCA has been used together with molecular quantum similarity measures (MQSM) to improve the identification of groups of molecules with similar characteristics \cite{fradera1997application,boon2001systematic}. 


\subsection{Astronomy}
The use and interest of PCA in astronomy has been growing in the last decades. Due to technological advances, new techniques for data capture and storage have been developed. In order to deal with this new data, PCA could be employed as an auxiliary tool. Applications range from analysis of data obtained from images~\cite{heyer1997application} to identification of stars~\cite{cabanac2002classification}, among other possibilities~\cite{deb2009light}. 

In another study, by employing PCA to a pulsar waterfall diagram, a method for determining the optimal periods of pulsar was proposed~\cite{nalettoapplication}. Observe that waterfall diagram means the data matrix obtained from the photon signal of the pulsar.

Conventional approaches for classification of astronomical objects and galaxy properties are based on artificial neural networks (ANNs)~\cite{storrie1992morphological,storrie1994spectral,folkes1996artificial,lahav1996neural,singh1998stellar,bailer1998automated}. ANNs are frequently used because of their non-linear classification capacity. In this context, some studies consider PCA as a complementary tool~\cite{lahav1996artificial,folkes1996artificial,lahav1996neural}. For instance, in the classification of astronomical objects, PCA was employed as a compression tool for the input data~\cite{lahav1996artificial}. Such an approach reduces the computational cost due to the lower amount of variables. Regarding stellar classification, other works used PCA to compress the stellar spectra, which is the data input of an ANN~\cite{singh1998stellar,bailer1998automated,storrie1994spectral}.  These studies indicate a relevant analysis of the compressibility of the stellar spectra. Additionally, PCA affected the classification accuracy, replicability, network stability, and convergence of the employed ANNs. 

Apart from the study considering neural networks, many other works employed PCA as a part of spectral analysis~\cite{paris2011principal,connolly1999robust,ronen1999principal,galaz1998eso,sodre1994spectral,sodre1997global}.
Taking into consideration the dimension reduction of PCA, \emph{Dultzin-Hacyan et al.}~\cite{dultzin1996general} studied the spectra provided from types 1 and 2 Seyfert galaxies. Interestingly, the spectrum of a type 1 Seyfert galaxy could be well described by a single component, but a type 2 Seyfert galaxy required at least three principal components. Still considering spectra information, PCA can also be used to improve classification schemes of galaxies~\cite{folkes19992df,connolly1995spectral}. Such methods analyze clusters in the PCA space, which are described by few principal components obtained from given galaxy spectra data.

\subsection{Geography}
Many works are related to geographical or spatial information~\cite{breschi2000geography,patterson2006population,cheng2013geographical,munoz2004spatio,demvsar2013principal,drafor2017principal}. For example, an exploratory study used PCA to analyse regularities in the distribution of innovative activities~\cite{breschi2000geography}. In other words, the authors investigated technological companies located in a same specific area. Data was obtained from patents of European countries, including France, Germany, Italy, and the UK. The information of the companies' addresses was used to infer their location, and PCA was applied subsequently.  In general, different results were obtained for distinct technological classes. In addition, for some of these classes, different countries presented similar trends. 

Another study related to geographical information is the investigation of the relationship between the genetic structure of human beings and their location~\cite{hofmanova2016early,novembre2008genes}. In~\cite{novembre2008genes}, information about genes of several European people was employed, and PCA was used to summarize the data. The geographic distance was found to influence the gene distribution.

Another work studied the determination of the procedence of food by employing chemical-related methods and PCA~\cite{kelly2005tracing,serra2005determination,rodrigues2009stable}. For example, studies analysed samples of green coffee, where PCA was used to determine where they were produced~\cite{kelly2005tracing,serra2005determination}. In both cases, the chemical characteristics of coffee were measured by using multivariate data obtained from Mass Spectrometry and visualized through projection onto the principal components.   

PCA can also be applied in order to study the geographical origin of propolis~\cite{cheng2013geographical}. Chemical experiments were performed, and the measured multivariate data was investigated by using PCA in the same fashion as in chemometrics (see Chemistry section).  By considering the three-dimensional PCA projection, four groups of samples were identified visually. 

An essential type of feature that can be used to describe regions is temporal information. A study about the rainfall patterns of Spain considered data from the years 1912 to 2000~\cite{munoz2004spatio}. In this work, PCA was employed as a preprocessing step of a clustering method to detect patterns of seasonal rainfalls. PCA was also applied as part of a work monitoring the growth of the urban area in Pearl River Delta region~\cite{li1998principal}.

\subsection{Weather}
Studies in meteorology and weather involve many different variables~\cite{daley1993atmospheric}. The high degree of freedom~\cite{daley1993atmospheric}, the chaotic behavior~\cite{rind1999complexity} and the difficulty of knowing the initial conditions~\cite{gneiting2005weather} of meteorological systems make the statistical approach attractive. PCA, in particular, is used in these areas both as part of forecast methods and as part of its data analysis~\cite{jaruszewicz2002application,sharma2011predicting}. Such studies are important for many applications, ranging from understanding the environment~\cite{maslennikova2015use} to practical applications~\cite{zarzo2011modeling,skittides2014wind}, such as the prediction of wind direction, essential for the high performance of wind energy generation~\cite{zarzo2011modeling,sharma2011predicting}. 

In order to improve the turbine performance of wind power stations, the prediction of wind speed and direction are essential parameters, and can vary substantially along time. Dealing with this problem, the authors of~\cite{skittides2014wind} proposed a forecasting method based mainly on PCA. The employed data are given by standardized time series of the wind speed and/or direction. By considering a subset of the wind measured data (the training set) and using \textit{Takens' method of delays} algorithm~\cite{takens1981detecting},  the delay matrix is computed. As a part of this method, PCA is applied to reduce the matrix dimensionality. The data of the test set (the remaining data) was converted to the same PCA space as the training set. Finally, by employing a strategy based on nearest neighbors, the current state of the wind is identified. Note that the neighborhood of the PCA  sample was used to predict its state and the forecast error.

Another study that aimed at better understanding weather characteristics in order to improve energy production in power stations was reported in~\cite{davo2016post}. PCA was used as a preprocessing tool for a forecasting method for wind and solar energy generation. As in the previous study, this technique was used to reduce the dimensionality of the input data, which were historical time series of power measurements of the power plants. The reduced data was employed as a training set for two different types of classifiers: (i)~Neural Network (NN) implemented by \emph{Venables and Ripley}~\cite{venables2013modern} and (ii)~Analog Ensemble (AnEn) algorithm~\cite{alessandrini2015novel}.

Other studies about climate employed PCA for finding Earth modifications. Examples include identifying consequences of climate change~\cite{loarie2008climate}, finding the source of pollutants~\cite{chan2007application}, and classifying bioclimatic zones~\cite{pineda2007regionalization}. Another study analyzed the ionospheric equatorial anomaly through a method called Total Electronic Content (TEC)~\cite{maslennikova2015use}. TEC is a descriptive method for Earth's ionosphere which counts the amount of electrons along a circular cross section of the atmosphere and integrates it between two points, giving the total number of electrons in that region.  PCA was used~\cite{maslennikova2015use} as a data analysis tool over TEC to aid the identification of temporal and spatial patterns.

\subsection{Agriculture}

PCA was used to study the origin of toxic elements in soils in~\cite{BORUVKA2005289}. While traditional techniques have been used for this purpose (profile and spatial distribution), it has been claimed that they are often unreliable to identify the sources of some elements in soils. Other traditional approaches such as parent rock decomposition and the knowledge of anthropogenic loads also yield inaccurate results with some frequency. 

The study conducted in~\cite{BORUVKA2005289} investigated the sources of pollution by analyzing the concentration of Cu, Hg, Ni, Pb, and Zn in the Czech Republic. It focused on the first three principal components, which accounted for 70\% of the data variance. A simple visualization allowed the identification of a cluster of elements, such as Co, Cr, Cu, Ni, and Zn. Interestingly, the data analysis showed that the main component could be interpreted as representing those elements of geogenic origin. Conversely, the third component was found to be able to identify pollution from atmospheric deposition. The results obtained in their study suggested that PCA can be employed as a tool to assist the identification of  the source of elements in soils.

\subsection{Tourism}
In Tourism, PCA has applications mainly on hotel location or recommendation systems. In~\cite{nilashi2015multi} a recommendation system based on collaborative filtering (data coming from similar users) was proposed. Here, PCA is used as a preprocessor to reduce the redundancy in the dataset. In another study~\cite{tkalcic2009personality}, PCA was used to assign weights when calculating the similarity between users of online tourism services.

The study proposed in \cite{hankinson2005destination} investigated the most important attributes of a given destination for a brand image in the framework of business tourism. It applied PCA as part of the \emph{varimax rotation} approach~\cite{coshall2000measurement} to the employed data set, which was acquired by an interview with different events managers. Such study suggests particular importance regarding the physical characteristics of a possible destination. Some examples of such type of attributes are: information about its architecture, number of attractive environments, and historical value.

\subsection{Arts}

PCA has been considered as a means to study  measurements obtained from pieces of art. There are two main types of analysis: assign scores to an art piece or composition~\cite{Luciano2012music}, or use their physical and/or chemical aspects~\cite{Luciano2015arts,Huang2012voice,Smaragdis2003polyphonic,Yang2012retagging,Baronti1997spectral,bacci1997principal}.

As an example of the first procedure, the researcher may establish variables, such as the presence of counterpoint and vocal melody lines in music~\cite{Luciano2012music}, or the presence of symbolism and contrasts in a painting~\cite{Luciano2015arts}. The corresponding feature space is then subjected to PCA in order to aid visualization and identify correlated features.  Other studies deal with the problem of determining the evoked emotions of music, in which a multiple linear regression (MLR) model is commonly used as a classifier. A large number of features related to the human perception can be extracted from a sound. Thus, in this framework, PCA is often used as a preprocessing tool for the reduction of the feature space to be fed into the MLR~\cite{yang2008regression,Chen2014music,Recours2009metal}. Furthermore, PCA can be used as an auxiliary methodology in other machine learning applications regarding music, such as music retrieval~\cite{whitman2001artist}, transcription~\cite{smaragdis2003non}, and genre classification~\cite{panagakis2009music}.

Other studies considered physical and chemical aspects. For instance, a composition's Fourier transform variation and sparsity, or a painting's reflection spectra, curvatures, areas, and ratios can be taken as the basis for the feature space~\cite{Luciano2015arts,Huang2012voice,Smaragdis2003polyphonic,Yang2012retagging,Baronti1997spectral}. In this framework, PCA was used as part of the supervised classification and separation algorithms for tagging and filtering music and art pieces. Image spectroscopy techniques were used to analyze paintings, and PCA was employed to reduce the dimensionality of the original data~\cite{baronti1998multispectral}. More specifically, oil paintings we analyzed and the proposed approach was able to characterize the pigments.  PCA was also employed as an auxiliary tool in the task of determining, in a non-invasive way, some chemical alterations in monuments~\cite{bacci1997principal}.


\subsection{History}

In History-related areas, and more specifically in the field of archeology and archaeological science, PCA is a well-established technique employed to help to analyze sets of prehistorical and historical objects and artifacts~\cite{baxter1994exploratory,wilcock1999getting}. In this approach, geometric measurements, chemical composition or other characteristics from the objects are understood as the features vector. PCA is then used to reduce the dimensionality, allowing the construction of bi-dimensional diagrams.

An example of the mentioned application is~\cite{vinagre2005ancient}, in which PCA is used to compare compositional data from ceramics between two archaeological sites from the same region in Brazil. The results, however, indicated no relationship between the communities that occupied these regions. In another study \cite{ravisankar2014energy}, the composition of 18th-century fragments collected in the Vellore Dist of Tamilnadu, India, were analyzed using PCA. The visualization of the principal components revealed the existence of three main clusters. Further investigations revealed that they correspond to different types of clay used to construct the objects. In a distinctive approach, PCA was employed as a tool to help to determine materials that could be used to restore monuments or historical buildings, while also maintaining similar properties under the same weather conditions~\cite{moropoulou2009principal}.

In~\cite{klinger2011landscape} PCA is used to help analyze and classify landscape features of regions surrounding archaeological sites. Particularly, geographical features, such as vegetation and landform characteristics are mapped to a lower dimensional space. With the help of a clustering technique, the data is compared with archaeological properties of the site, which includes information such as the existence of a settlement or the type of buildings. The analysis revealed the existence of patterns that could be used to identify other, yet to be discovered, archaeological sites.

Another use of PCA in History is to track and understand the dynamics of populations according to genetic changes and mixture. In particular, there is interest in finding events of genetic admixture, in which two previously isolated populations start to interact and grow~\cite{templeton2006population}. An example of such an approach is \cite{chunjie2000principal}, in which the distribution of a certain set of genetic markers among different Chinese populations is used as input for PCA. The most informative components are then projected over the maps of the corresponding geographical regions for each population. The obtained visualizations allowed historians to track the migration patterns of minorities along the history of China. In \cite{moorjani2011history}, a more complex approach was employed to track the African ancestry among populations of many regions. In a different approach~\cite{chen1983surnames}, surnames distribution are used instead of genetic markers, tracking the history of Taiwanese populations using tree-based clustering and PCA. 

\subsection{Social Sciences}
Social sciences include many aspects of different disciplines that are related to individuals and society. Some examples of subjects are economics, geography, history, and linguistics. Here, we describe an overview of PCA applied as a tool to address problems in social sciences. 

In economics, many indices are measured and then analyzed. 
PCA has been used to investigate socio-economic status indices~\cite{vyas2006constructing}. Other studies about economics have also employed PCA. Multivariate temporal data of many economic indicators from India were evaluated by using PCA in order to study financial development of that country~\cite{lenka2015measuring}. In this way, an index named FDI (Financial Depth Index) was proposed. FDI is computed by applying PCA to the indicators mentioned above. So, the multivariate data could be summarized by a single index. 

Apart from using numerical variables, some information regarding socioeconomics is related to categorical variables~\cite{kolenikov2009socioeconomic}. Modifications of PCA were proposed by considering several different ways to transform the binary categorical variables into continuous ones. One way to treat this type of data is to apply techniques that transform categorical data into dummy variables~\cite{filmer2001estimating}, by employing the Filmer–Pritchett procedure. However, this alternative should be used only when the order of the information is unknown~\cite{kolenikov2009socioeconomic}. Otherwise, other techniques, such as the ordinal PCA, are more trustworthy~\cite{kolenikov2009socioeconomic}.

Another aspect studied in social sciences is the structure of social networks. In a study analysing a web-based social network, PCA was applied as a statistical tool to find patterns in the network according to cultural interests~\cite{liu2007social}. As another example, PCA was employed as an auxiliary technique of factor analysis to better understand attraction preferences~\cite{gangestad2004women}.

\subsection{Linguistics}
In text analysis, a popular application of principal component analysis is in stylometry. In this context, the PCA technique is often referred to as eigenanalysis of function words, since the most traditional techniques in textual style identification concern the use of function words such as articles and prepositions~\cite{doi:10.1093/llc/14.4.445,AMANCIO20124406}. In~\cite{doi:10.1093/llc/14.4.445}, an intuitive introduction of PCA is presented, with application to a structural comparison of written documents. In talks about distinguishing between authors, a statistical test is applied in order to identify the most discriminating words in the problem of distinguishing Shakespeare's and Wilkins' works. Among the most discriminating words found we have: \emph{then}, \emph{the} and \emph{by}. The PCA transformation considering the original set of 20 features yielded a compression rate of 54\% and 15\%, respectively for the first and second principal components. The projection of the textual data revealed that the work treated as of unknown authorship (Winter's Table) in the study is more similar to Shakespeare's work. By using the same dataset, the authors also found that ``\emph{The Tempest}'', which is known for its controversial hybrid style, is surprisingly consistent with the style developed by Shakespeare, according to its projection onto the first principal component. Although PCA is not used as a discrimination method, the authors suggest that this dimensionality reduction method can shed light on the problem of quantifying the similarity between literary works.  

In addition to the traditional studies in stylometry relying on the frequency of function words, novel features have been proposed more recently. The network-based approach proposed in~\cite{1367-2630-14-4-043029} relies on a small number of features to categorize texts using analysis of components. The authors represent text as a complex network, where words are nodes and edges are adjacency relationships. In order to characterize the documents, network metrics such as betweenness and clustering coefficient are extracted. The obtained multivariate space is then compressed in two main components, with a high compaction degree. The high value of compaction is in accordance with the several correlations found in network-structured data representing documents and other real systems. Interestingly, the first two principal components were able to discriminate between literary movements according to the visualizations generated by reducing the dimensionality of the original data. While the authors used the reduction of dimensionality as a visualization tool, the same methodology could be adapted as a pre-processing step in classifying documents. Similar studies combining network science, stylometry, text analysis and principal component analysis can be found in~\cite{1742-5468-2015-3-P03005,2015comparing,de2016using,marinho2017Calligraphy,7033961,10.1371/journal.pone.0118394}.

The PCA technique has also been employed in linguistic studies devoted to assisting diagnosis of some diseases. Particular examples are the studies applied to identify linguistic variations in Alzheimer's patients~\cite{MARCIE199353}. Actually, it has been long known that Alzheimer's disease affects linguistic features in a way similar to aphasia~\cite{Lopez:2011:PCA:1959868.1959935}. Based on such evidence, the study conducted in~\cite{MARCIE199353} evaluated how the disease influences language in 15 different features to analyze the ability to generate sentences from some given words, the ability to name objects, and identify antonyms. Each feature was analyzed regarding the error rate generated by the patients. A principal component analysis of the considered features was found to be efficient to account for most of the observed variance by considering the two principal components. The study showed that the first principal component aggregated mainly the features related to oral language. Conversely, the second component aggregated aspect related to written language, which included both writing and reading skills. Though this study did not focus on a more systematic analysis to visualize and classify data, the PCA analysis was helpful to show that the most substantial variation of the studied features in patients is their ability to generate antonyms. 

In addition to being useful to visualize data obtained from language applications, the PCA can alternatively be used as a pre-processing step in text classification tasks. In the study carried out in~\cite{UGUZ20111024}, the authors aimed at classifying documents in several categories with a low computational cost. In order to perform the classification, the proposed method removed words conveying low semantic meaning (i.e., the \emph{stopwords}). Then, the remaining words were stemmized and tf-idf (term frequency-inverse document frequency) weighting value~\cite{Manning:1999} was assigned to each word.  By using concepts from information theory, the best features (i.e., the most discriminative words) were selected according to the mutual information metric. After such a feature selection, PCA was applied to the remaining features as an additional feature selection step. As a criterion to select an adequate number of features, the method selected the principal components so as to keep 75\% of the original data variability. The classification algorithm in the compressed data revealed that a high accuracy rate can be obtained even if several features are disregarded. The authors reported that when the total number of considered features decreases from 319 to 75, the accuracy only drops 3\% in the textual classification task.

\section{Experimental Study of PCA Applied to Diverse Databases}

One of the main reasons justifying the popularity of PCA is its ability to reduce the dimensionality of the original data while preserving its variation as much as possible. In addition to allowing faster and simpler computational analysis, these reduction also allows the identification of new important variables with distinctive explanatory capabilities.  As reported in the literature, a small number of PCA variables often can account for most of the variance explanation.  In this section we develop an experimental approach aimed at quantifying in more objective terms the explanatory ability of PCA with respect to some representative real-world databases derived from the previously surveyed material. Here, we considered two datasets representative of each of 10 different areas. The selected areas are astronomy, biology, chemistry, computer science, engineering, geography, linguistics, materials, medical, and weather. After pre-processing required for eliminating incomplete and categorical data, PCA was applied to the databases. Two cases were considered for generality's sake: (a) without standardization; and (b) with standardization.  The possible effects of the database size, number of features, and number of categories on the variance explanation were also investigated.

\subsection{Dataset Selection}

For all considered datasets, we eliminated non-numerical data, such as categorical values and dates. Some characteristics of the considered datasets are shown in Table~\ref{tab:pcaDatabase}.

In the astronomy area, we considered data from \emph{galaxies}~\cite{willett2013galaxy}. More specifically, a table that comprises measurements of spectroscopic redshifts.  The second dataset comprised \emph{ionosphere} as measured by radar~\cite{sigillito1989classification}. In biology, we considered datasets regarding \emph{gene} expression levels~\cite{eisen1998cluster} and measurements of \emph{leaves}~\cite{silva2013evaluation,Dua:2017}. Two datasets of food were employed representing chemistry: (i) data on characteristics of \emph{wine}~\cite{forma1988parvus,Dua:2017} and (ii) data of \emph{milk} composition~\cite{daudin1988stability}. In case of computer science, two different subjects were considered, which are characteristics of computers~\cite{Dua:2017} (\emph{machine}) and features of image \emph{segmentation} data. This image segmentation (\emph{segment-challenge}) dataset is part of the datasets provided by the software Weka~\cite{witten2016data}. In engineering, we used a dataset of the electric power consumption in houses~\cite{Dua:2017} (\emph{energy}) and information of concrete \emph{slump} tests~\cite{yeh2006exploring,Dua:2017}. 

The datasets of geography contain data of spatial coordinates and information related to weather. The first dataset is about \emph{dengue} disease~\cite{hales2002potential} and the second is about \emph{forest} fires~\cite{cortez2007data,Dua:2017}. In linguistics, the first dataset comprises the frequency of linguistic elements (eg., punctuation and symbols) in texts of commerce \emph{reviews}~\cite{liu2011application,Dua:2017}; the second one contains statistics of \emph{blog} feedbacks~\cite{buza2014feedback,Dua:2017}. The datasets considered in materials area are \emph{glass} identification~\cite{evett1987rule}, with information of refractive index and chemical elements, and measurements regarding \emph{plates} faults~\cite{platesDataset,Dua:2017}. In medical, the used dataset is about characteristics of people with or without \emph{diabetes}~\cite{smith1988using}, and the other dataset considers biomedical voice measurements regarding patients with or without \emph{Parkinson} disease~\cite{little2007exploiting,Dua:2017}. Finally, the datasets of weather are: (i) environmental measures of 
\emph{El Ni\~no}~\cite{bay2000archive,Dua:2017} and (ii) measures related to \emph{ozone} level~\cite{Dua:2017}.

\begin{table}[!h]
\centering
\begin{tabular}{lccc}
\hline
\hline
Name & Classes & Samples & Measurements \\ \hline
astronomy (galaxy) & - & 243,500 & 225\\ 
astronomy (ionosphere) & 2 & 351 & 34\\ 
biology (gene) & 6 & 545 & 78\\ 
biology (leaf) & 36 & 340 & 14\\ 
chemistry (milk) & - & 86 & 7\\ 
chemistry (wine) & 3 & 178 & 13\\ 
computer (machine) & - & 209 & 7\\ 
computer (segment-challenge) & 7 & 1500 & 19\\ 
engineering (energy) & - & 2,049,280 & 6\\ 
engineering (slump) & - & 103 & 9\\ 
geography (dengue) & 2 & 1,986 & 11\\ 
geography (forest) & - & 517 & 10\\ 
linguistics (reviews) & 50 & 1,500 & 10000\\ 
linguistics (blog) & - & 52,397 & 280\\
materials (glass) & 7 & 214 & 9\\ 
materials (plates) & 7 & 1,941 & 26\\ 
medical (diabetes) & 2 & 768 & 8\\ 
medical (parkinsons) & 2 & 195 & 21\\ 
weather (el ni\~no) & - & 533 & 6\\ 
weather (ozone) & - & 1,847 & 71\\ 
\hline
\hline
\end{tabular}
\centering
\caption{Characteristics of the considered databases, concerning different areas.}
\label{tab:pcaDatabase}
\end{table}

\subsection{Results and Discussion}

\begin{figure}[ht!]
    \centering
    \subfigure[Standardized.]{\includegraphics[width=0.49\textwidth]{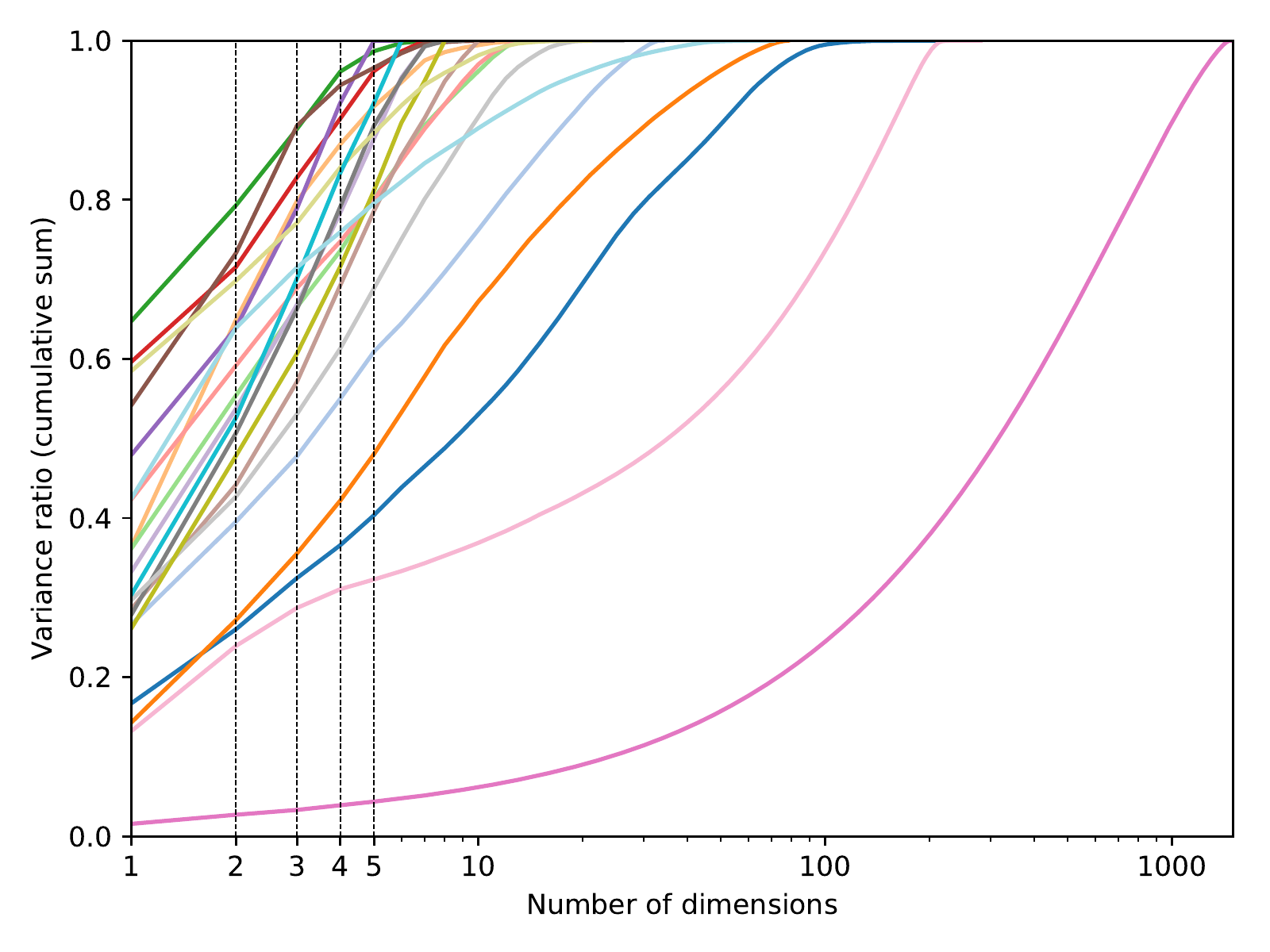}}
    \subfigure[Without standardization.]{\includegraphics[width=0.49\textwidth]{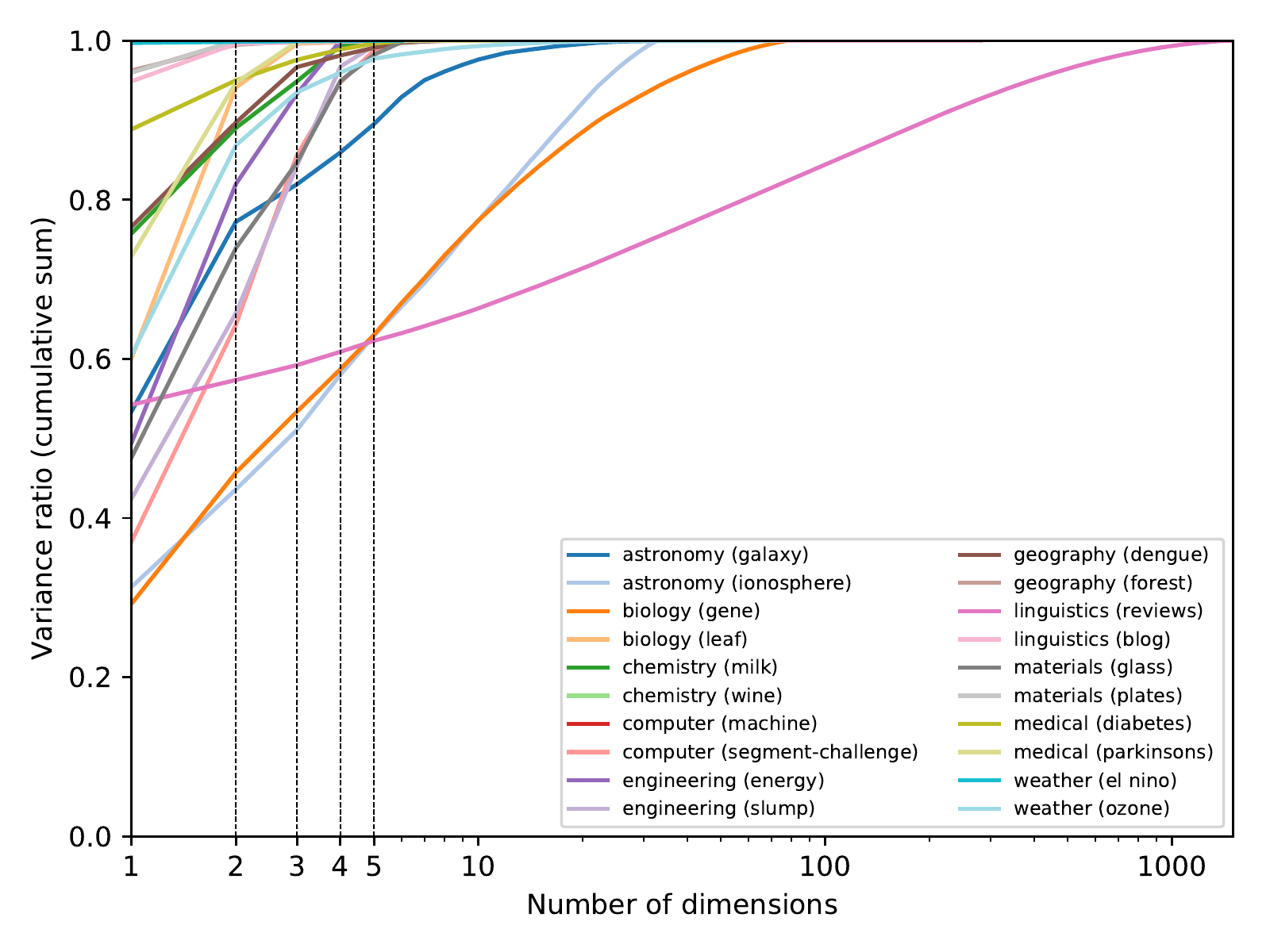}}
    \caption{PCA variance ratio according to the number of principal components, for all the datasets presented in Table~\ref{tab:pcaDatabase}. The vertical dashed lines indicate the situations where the first 2, 3, 4 and 5 PCA components are retained.}\label{fig:logcompac2}
\end{figure}

In order to compare the amount of variance retained by PCA for the different datasets, in Figure~\ref{fig:logcompac2} we plot the number of PCA components against the respective variance ratio, defined by Equation~\ref{eq:PCA_var_ratio}. Figure~\ref{fig:logcompac2}~(a)~and~(b) show the measurements for standardized and non-standardized data, respectively. Note that in the majority of the cases for standardized data, the first three principal components can represent more than 50\% of the variance in the datasets. By considering the data without standardization, 60\% of the variance is contained in the first two principal components on the majority of the datasets. This is because, when the data is not standardized, a few measurements having large values dominate the variance in the data. One example of such effect is shown in the linguistics (reviews) dataset, which have more than 50\% of its variance explained by a single component without standardization, but negligible variance explanation for a single component if the data is standardized.

\begin{figure}[ht!]
    \centering
    \subfigure[astronomy (ionosphere)]{\includegraphics[width=0.22\textwidth]{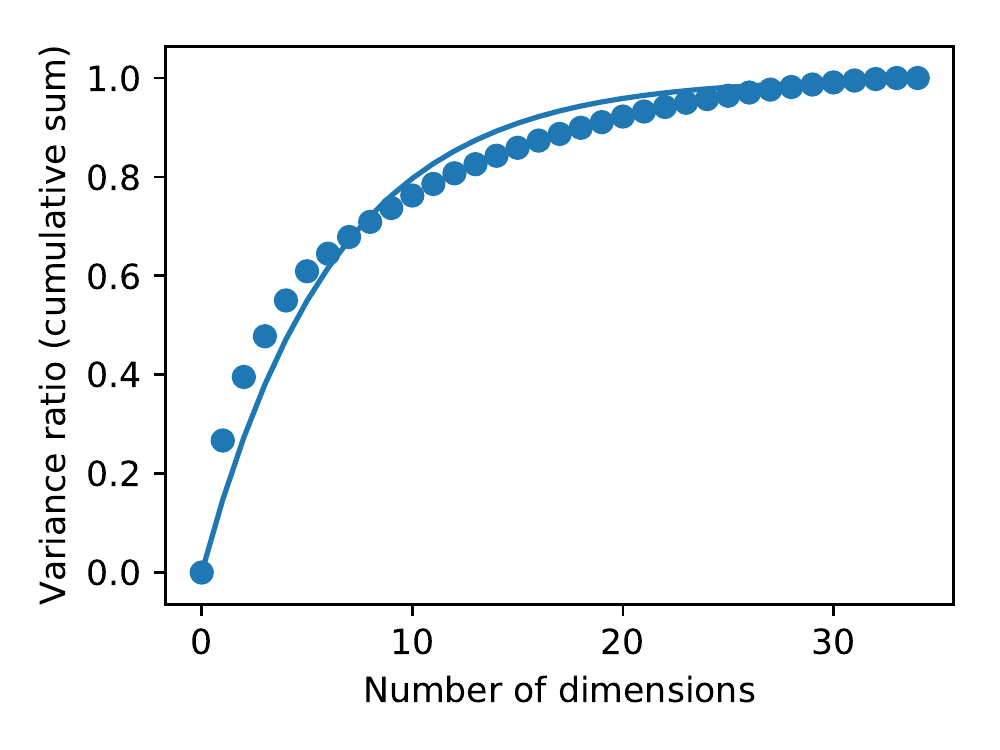}}
    \subfigure[engineering (slump)]{\includegraphics[width=0.22\textwidth]{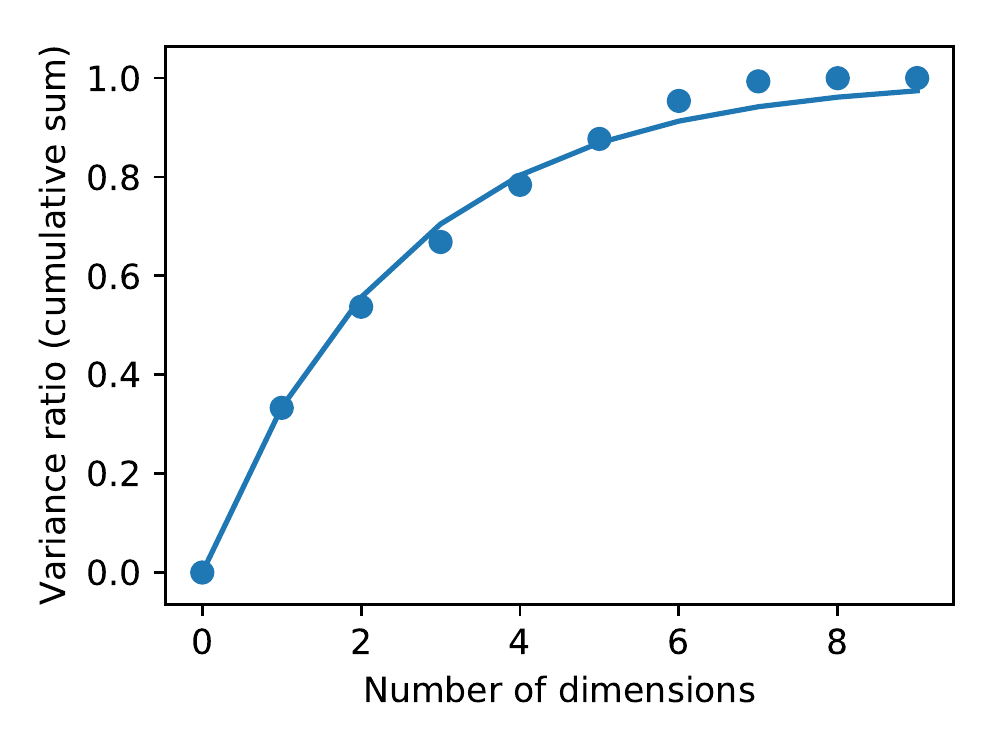}}
    \subfigure[geography (dengue)]{\includegraphics[width=0.22\textwidth]{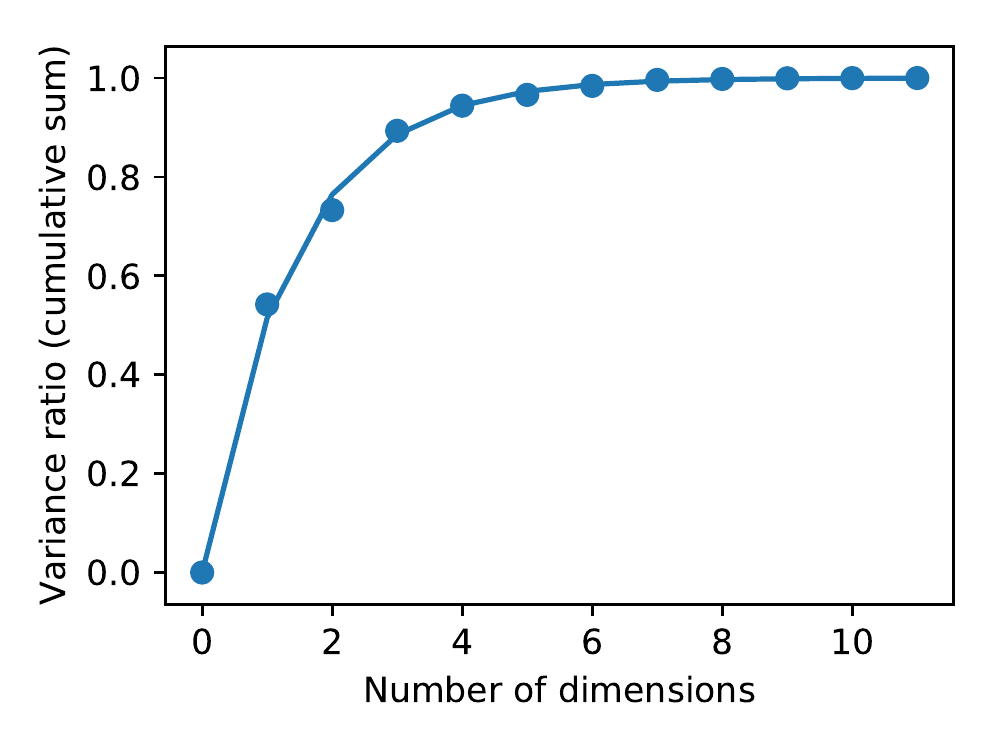}}
    \subfigure[materials (plates)]{\includegraphics[width=0.22\textwidth]{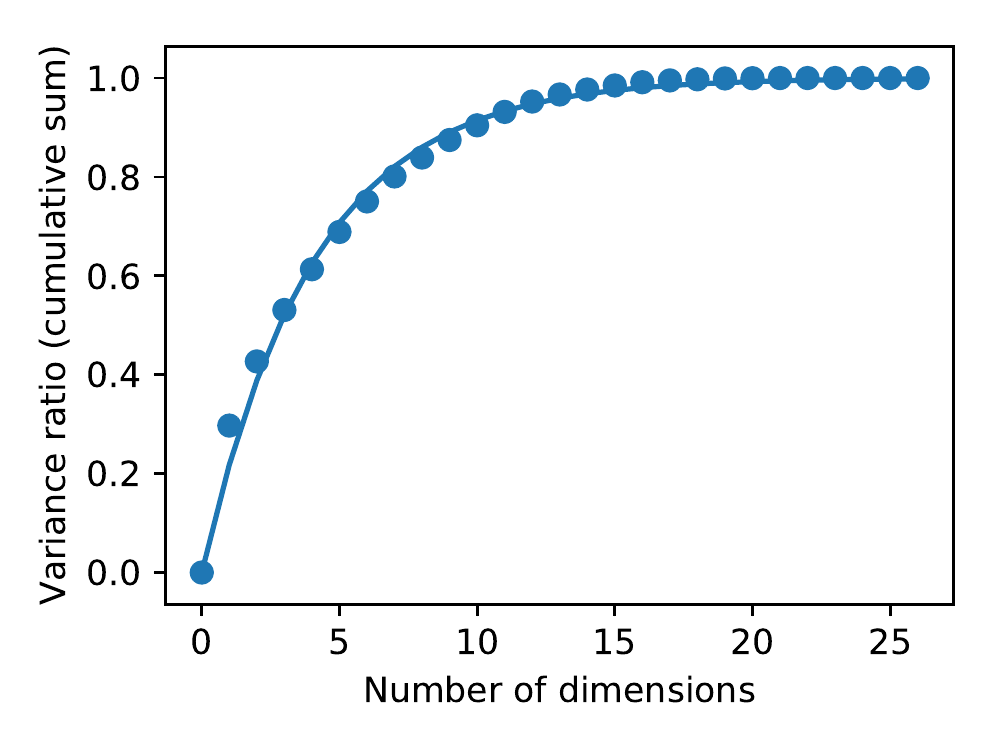}} 
    \caption{Examples of the curve fits of the cumulative sum of the variance ratio against the number of dimensions, for standardized data. The circles represent the measured data ant the line is the fitted curve.}\label{fig:fits}
\end{figure}

In case of the standardized data, the majority of the curves of accumulated variance ratio seems to follow the function
\begin{equation}
f(x) = 1 - \exp{(-\alpha x)},
\end{equation}
where $\alpha$ is a parameter. So, we employed the \emph{Levenberg-Marquardt} algorithm~\cite{more1978levenberg}, known as damped least squares, to find the value of $\alpha$ to fit each curve. In order to visualize the adjusted functions, we plot, for some datasets, the original data and the respectively fitted curve in linear scale, as shown in Figure~\ref{fig:fits}. As a complementary analysis, we compute the Pearson correlation between the obtained $\alpha$ values and characteristics of the datasets. For the properties number of classes, number of samples, and number of measurements, the measured correlation values are, respectively, -0.27, 0.15, and -0.38. Low values of Pearson correlation were also found for the non-standardized data. This result indicates that there is no strong correlations between the characteristics of the considered datasets and $\alpha$. Therefore, the amount of variance retained by the PCA axis cannot be explained by these properties alone.

\begin{figure}[ht!]
    \centering
	\includegraphics[width=0.49\textwidth]{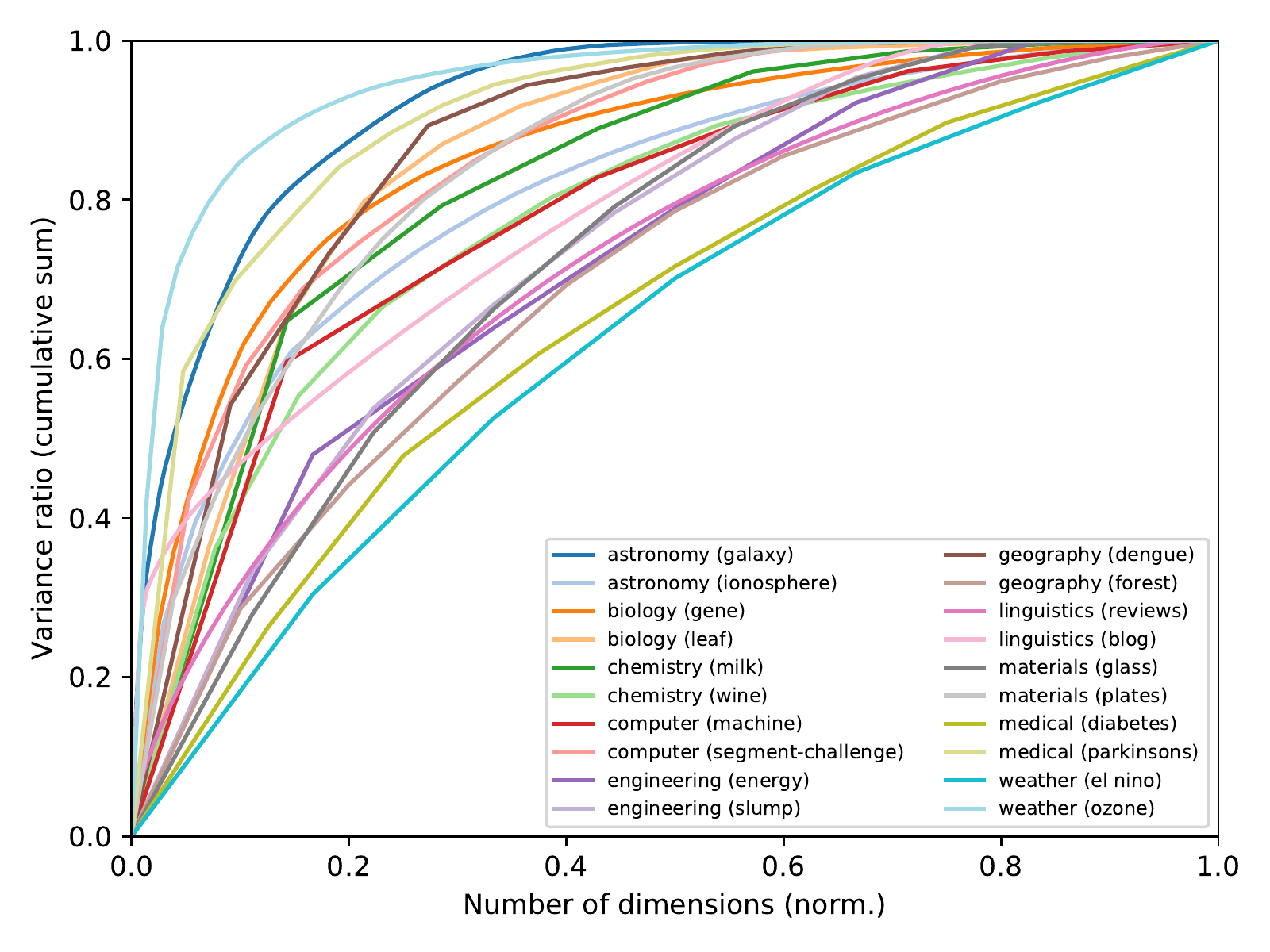}
    \caption{PCA variance ratio against normalized number of dimensions, for all the datasets presented in Table~\ref{tab:pcaDatabase}. Note that the normalized number of principal components consists in the fraction of components computed for each dataset. In this experiment all the data were standardized.}\label{fig:compac1}
\end{figure}

As a complementary analysis, we compare the variance ratio among the datasets by normalizing by the number of dimensions in the original dataset (see Figure~\ref{fig:compac1}). The main purpose of this analysis is to identify the number of PCA components that need to be retained, compared to the original number of features in the dataset, in order to obtain specific variance ratio values. By comparing Figures~\ref{fig:compac1} and~\ref{fig:logcompac2}~(a), we note that the relative order of the curves change. This means that some datasets need many PCA components in order to achieve large variance ratio, but since these datasets originally have a large number of features, only a small percentage of the number of features is enough to represent most of the variance in the data. Note that the variance explanation is lower for curves near the diagonal. The dataset having the nearest curve to the diagonal is \emph{weather (el nino)}. 

\begin{figure}[ht!]
    \centering
   \includegraphics[width=0.49\textwidth]{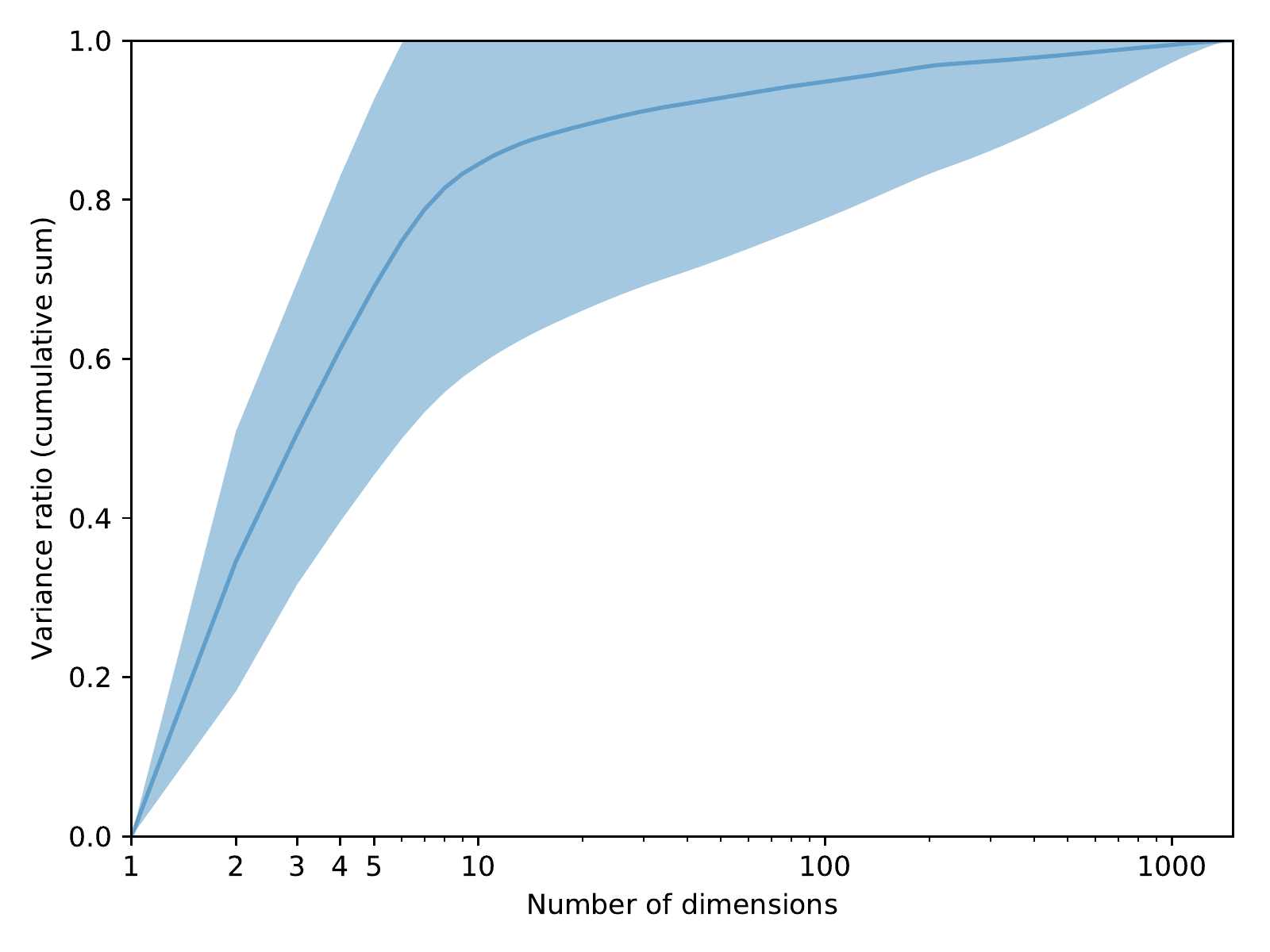}
    \caption{By considering the standardized datasets, the curves are the average PCA variance ratio against the number of dimensions (in log scale), for all the datasets presented in Table~\ref{tab:pcaDatabase}. The shaded area represents the standard deviation.}\label{fig:avg}
\end{figure}

\begin{figure}[ht!]
    \centering
    \includegraphics[width=0.49\textwidth]{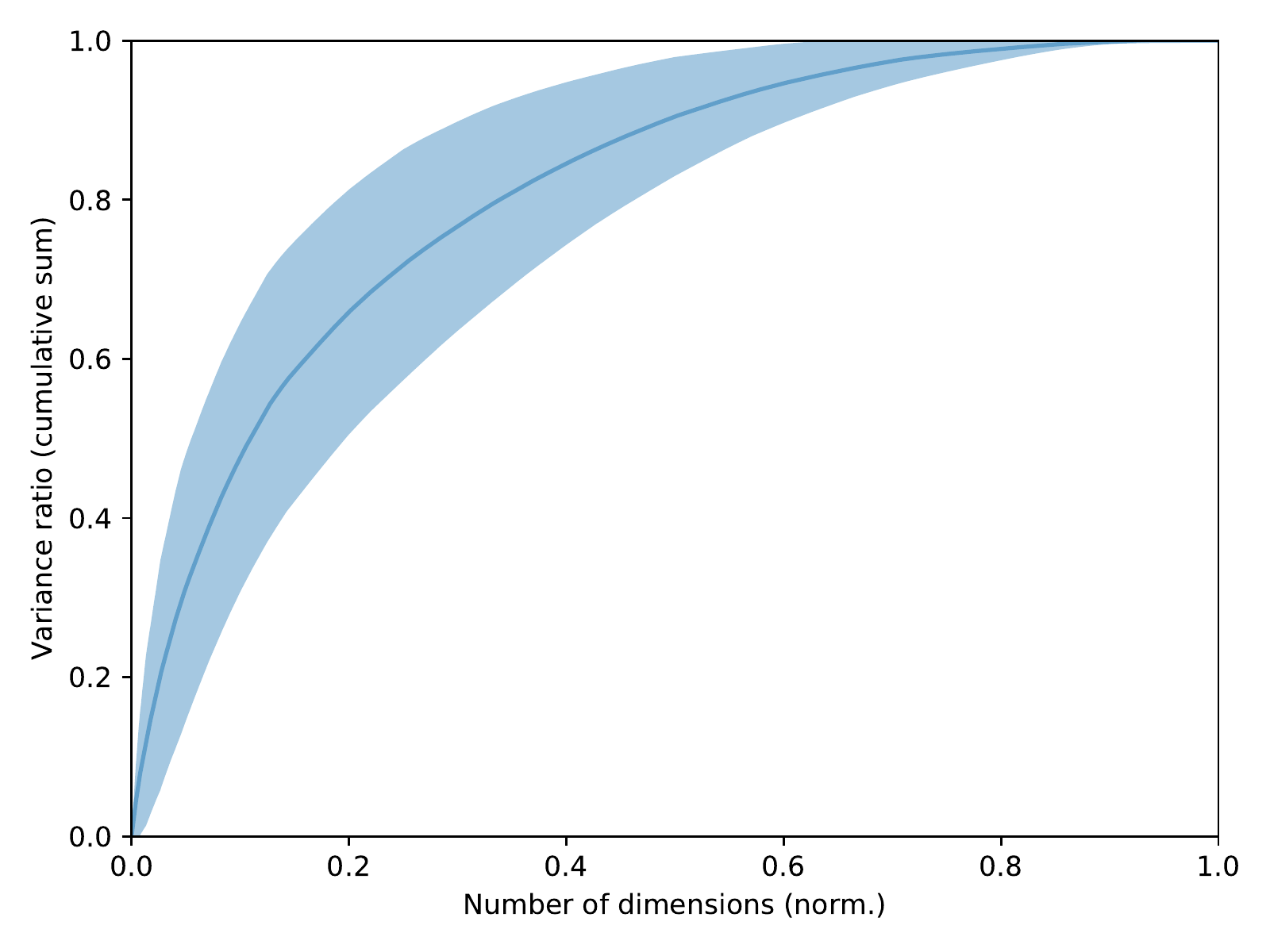}
    \caption{In the case of the standardized datasets, the curves correspond to the average PCA variance ratio against the normalized number of dimensions, for all the datasets presented in Table~\ref{tab:pcaDatabase}. The shaded area represents the standard deviation.}\label{fig:avgNormed}
\end{figure}

In order to summarize the results, we plot the average curve, taken over all datasets, of the PCA variance ratio as a function of the number of principal components. Figure~\ref{fig:avg} shows the typical curve obtained without normalizing by the number of features in the dataset, while Figure~\ref{fig:avgNormed} shows the average curve when applying the normalization. In both cases, the data was standardized before applying PCA.

\section{Concluding Remarks}

Principal component analysis -- PCA -- has become a standard 
approach in data analysis as a consequence of its ability to 
reduce dimensionality while preserving variance of the data.  
In this work, we reported an integrated and systematic review 
of PCA covering several of its theoretical and applied aspects.
We start by providing a simple and yet complete application 
example of PCA to real-world data (beans), and by identifying
three typical ways in which PCA can be applied.
Next, we developed the concept of PCA from more basic aspects
of multivariate statistics, and present the important issues
of variance preservation.  The option to normalize or not the
original data is addressed subsequently, and it is shown that
each of these alternatives can have major impact on the obtained
results.  Guidelines are provided that can help the reader to
decide wether to normalize or not the data.  Other
aspects of PCA application are also addressed, including the 
direction of the principal axes, the relationship between
PCA and rotation, and the demonstration of maximum variance
of the first PCA axes.  Another projection approach, namely
LDA, is briefly presented next.

After presenting the several aspects and properties of PCA,
we develop a systematic, but not exhaustive, review of some
representative works from several distinct areas that have
used PCA for the most diverse applications.  This review
fully substantiates the generality and efficacy of PCA for
a wide range of data analysis applications, confirming its
role as a choice method for that finality.

The last part of this work presents an experimental investigation
of the potential of PCA for variance explanation and dimensionality
reduction. Several real-world databases are considered, founded
on the main areas reviews in the previous sections. The obtained
results confirm the ability of PCA for explaining several types
of data while using only a few principal axes. Special attention
was given to the study of the effects of data standardization on
variance explanation, and we found that non-standardized data
tend to yield more intense variance explanation. We also showed
that the variance ratio curves can be reasonably well fitted by
using the exponential function. This result allowed us to 
quantify the effect of data size, number of classes and number
of features on the overall variance explanation. Interestingly,
it has been found that these properties do not tend to have
any pronounced influence.

All in all, we hope that the reported work on PCA covering
from basic principles to a systematic survey of applications,
and including experimental investigations of variance explanation,
can provide resources for researchers from the most varied areas
that can help them to better apply PCA and interpret the respective
results.

\section*{Acknowledgments}

Gustavo R. Ferreira acknowledges financial support from CNPq (grant no. 158128/2017-6). Henrique F. de Arruda acknowledges CAPES for sponsorship. Filipi N. Silva thanks FAPESP (grant no. 2015/08003-4 and 2017/09280-7) for sponsorship. Cesar H. Comin thanks FAPESP (grant no. 15/18942-8) for financial support. Diego R. Amancio acknowledges financial support from FAPESP (16/19069-9 and 17/13464-6). Luciano da F. Costa thanks CNPq (grant no. 307333/2013-2) and NAP-PRP-USP for sponsorship. This work has been supported also by FAPESP grants 11/50761-2 and 2015/22308-2.

\section*{Appendix A - Symbols}

\begin{table}[!h]
\begin{tabular}{ll}
  \hline
  \hline
	Number of features  & $N$ \\
    Number of projected features & $M$ \\
    Number of objects & $Q$ \\
    Data matrix & $X$ \\
    $i$-th feature & $X_i$ \\    
    $i$-th feature for all objects & $\vec{X}_i$ \\
    $i$-th feature of object $j$ & $X_{ij}$\\
    Average vector of data matrix $X$ & $\vec{\mu}_X$ \\
    Average of feature $X_i$ & $\mu_{X_i}$ \\
    Standard deviation vector of data matrix $X$ & $\vec{\sigma}_X$ \\
    Standard deviation of feature $X_i$ & $\sigma_{X_i}$    \\
    Transformed data matrix & $Y$ \\
    $i$-th transformed feature & $Y_i$ \\
    $i$-th transformed feature for all objects & $\vec{Y}_i$ \\ 
    Feature vector of object $i$ & $\vec{\mathscr{X}}_i$ \\
    Transformed feature vector of object $i$ & $\vec{\mathscr{Y}}_i$ \\
    PCA transformation matrix & $W$ \\
    $i$-th row of $W$ & $\vec{v}_i$ \\
    Correlation matrix & $R$ \\
    Covariance matrix & $K$ \\
    Pearson correlation matrix & $C$ \\
    Pearson correlation between the $i$th and $j$th variables & $C_{ij}$ \\
    Pearson correlation between variables $X_i$ and $X_j$ & $\rho_{X_i X_j}$ \\
    Percentage of preserved variance & $G$ \\
    Expectation of random variable $A$ & $E[A]$ \\
  \hline
  \hline
\end{tabular}
\caption{\label{t:dataOrganize} Description of the main symbols used in this work.}
\end{table}

\section*{Appendix B - Consequences of Normality}






As shown in the main text, PCA returns a maximal variance projection for any dataset of finite variance. If the underlying data follow a normal (Gaussian) distribution, more can be said: as we show here, the principal components are independent and maximize the projection's entropy given the original data.

Firstly, we recall that an $N$-dimensional random variable $\vec{X}$ follows a normal distribution with mean $\vec{\mu}$ and covariance matrix $\Sigma$ -- denoted $X \sim \mathcal{N}(\vec{\mu}, \Sigma)$ -- if its probability density function is expressed as
\begin{equation}
f(\vec{x}) = \frac{1}{(2\pi)^{N/2}\lvert\Sigma\rvert^{1/2}}e^{-\frac{1}{2}(\vec{x} - \vec{\mu})^T\Sigma^{-1}(\vec{x} - \vec{\mu})}, \end{equation}
where $\lvert\Sigma\rvert$ denotes the determinant of $\Sigma$. We also state some basic properties of a normal distribution:

\begin{lem}
\label{lem:1}
Let $\vec{X} = (X_1, X_2, \ldots, X_N)$ be a random variable following a Gaussian distribution with mean $\vec{\mu}$ and covariance matrix $\Sigma$, then:
\begin{enumerate}[(i)]
\item Its components $X_1, \ldots, X_N$ are also normally distributed.
\item Let $A$ be a real matrix with $N$ columns. Then, $\vec{Y} = A\vec{X}$ follows a normal distribution with parameters $A\vec{\mu}$ and $A\Sigma A^T$.
\item Two components $X_i$ and $X_j$, $1 \leq i,j \leq N$ are independent if and only if they are uncorrelated.
\end{enumerate}
\end{lem}

For a proof, see Anderson~\cite{Anderson2003multi}, section 2.4. Now, we turn these properties to the context of PCA. Recall that the principal components correspond to transformations of the coordinate axes under matrix $W$. Thus, if the original dataset $X$ is normally distributed with covariance matrix $\Sigma$, the transformed data is $Y = WX$, and we conclude from Lemma \ref{lem:1} that $Y$ follows a Gaussian distribution with covariance matrix $Cov(Y) = W\Sigma W^T$. From our previous discussions (see Section \ref{s:decorrDemonst}), we obtain immediately that $Cov(Y)$ is a diagonal matrix. Therefore, its components (which correspond to the principal components of our dataset) are independent on top of being uncorrelated.

Next, we turn to information-theoretic properties of a normal distribution and its projections. As defined by Shannon, the entropy of an $N$-dimensional random variable $\vec{X}$ with probability density function $p(\vec{x})$ is given by~\cite{Thomas2006info}
\begin{equation}
H[\vec{X}] = -\int_{\mathbb{R}^N} p(\vec{x})\ln p(\vec{x})\,d\vec{x}.
\end{equation}

Informally, entropy is a measure of the distribution's uncertainty, or lack of information. Consequently, maximizing the entropy of a distribution is equivalent to avoiding imposing additional hypotheses and constraints on data (see, for example, Jaynes~\cite{Jaynes1957MaxEnt} and Caticha~\cite{Caticha2011AIP,Caticha2004Bayes}). Therefore, finding a maximal entropy projection of a dataset is a problem of great interest in exploratory analyses and dimensionality reduction.

Now, consider the problem of maximizing a distribution's entropy subjected to certain constraints. Suppose we want to obtain a distribution $p(x)$ satisfying $E[X] = \mu$ and $E[X^2] = \sigma^2$. Then, the Lagrange multiplier theorem tells us that the solution is a minimum of the functional 
\begin{multline}
J[p] = -\int_{-\infty}^{+\infty} p(x)\ln p(x)\,dx + \\ + \lambda_0\left[1 - \int_{-\infty}^{+\infty} p(x)\,dx\right] + \\ + \lambda_1\left[\mu - \int_{-\infty}^{+\infty} xp(x)\,dx\right] + \\ + \lambda_2\left[\sigma^2 - \int_{-\infty}^{+\infty} x^2p(x)\,dx\right].
\end{multline}

Differentiating with respect to $p$ (in the context of the calculus of variations; see, for instance, \cite{Gelfand1963cv}) and equating to zero, we obtain:

\begin{equation}
-\left(\ln p(x) + 1 + \lambda_0 + x\lambda_1 + x^2\lambda_2^2\right) = 0.
\end{equation}

We solve this equation for $p(x) = e^{- 1 -\lambda_0 - x\lambda_1 - x^2\lambda_2^2}$. Since the exponent is a quadratic form, we recognize it as a normal distribution, and we set the values of the $\lambda_i$'s by imposing the normalization and moment constraints. We conclude that $X \sim \mathcal{N}(\mu, \sigma^2 - \mu^2)$ is a maximum entropy distribution for given mean and variance (recall that $Var(X) = E[X^2] - E[X]^2$).

An analogous calculation for a multivariate case (see \cite{Thomas2006info}) shows that the multivariate normal distribution is also a maximum entropy distribution for a given mean vector $\vec{\mu}$ and covariance matrix $\Sigma$. Its entropy can be analytically expressed as~\cite{Thomas2006info}
\begin{equation}
H[\vec{X}] = \frac{1}{2}\ln \lvert 2\pi e\Sigma\rvert.
\end{equation}

We see that it depends only on the covariance matrix's determinant. Thus, a first conclusion is that the PCA preserves the distribution's entropy; since the determinant is invariant under a transformation of the form $W\Sigma W^{-1}$, it follows that $\lvert Cov(\vec{Y}) \rvert = \lvert\Sigma\rvert$ and thus $H[\vec{Y}] = H[\vec{X}]$.

But usually PCA consists of taking a submatrix of the transformed dataset $Y$. The covariance matrix corresponding to these data is a leading principal submatrix of the diagonal matrix $Cov(Y) = W\Sigma W^T$, corresponding to its largest eigenvalues. Since a matrix's determinant is the product of its eigenvalues, taking the principal minor with the largest eigenvalues as done in the PCA gives us a maximum entropy projection; recall that the projected data are also normally distributed, and a normal distribution's entropy is monotonically increasing in the covariance matrix's determinant.

\section*{Appendix C - Biplot background }
\label{s:biplot_calc}

The Pearson correlation coefficient between standardized variable $\tilde{X}_j$ and PCA component $Y_i$, calculated from matrix $\tilde{X}$, is given by

\begin{align}
PCorr(Y_i, \tilde{X}_j) & = \frac{1}{Q-1}\sum_k \frac{Y_{ik}\tilde{X}_{jk}}{\sigma_{Y_i}} \\
& = \frac{1}{(Q-1)\sqrt{\lambda_i}}\sum_k Y_{ik}\tilde{X}_{jk},\label{eq:pcorrPCAC}
\end{align}
where $Q$ is the number of objects. The $k$th value of PCA component $i$ is a linear combination of the original measurements weighted by the respective eigenvector, that is

\begin{equation}
Y_{ik} = \sum_l W_{il}\tilde{X}_{lk}\label{eq:kPCAcomp}
\end{equation}
Replacing Equation~\ref{eq:kPCAcomp} in Equation~\ref{eq:pcorrPCAC}, we obtain

\begin{equation}
PCorr(Y_i, \tilde{X}_j) = \frac{1}{(Q-1)\sqrt{\lambda_i}}\sum_k\sum_l W_{il}\tilde{X}_{lk}\tilde{X}_{jk} \label{eq:corr_pcai_xj}
\end{equation}

Since $\vec{v}_i=(W_{i1},\dots,W_{iN})$ is an eigenvector of the correlation matrix C, we have that

\begin{equation}
C\vec{v}_i = \lambda_i \vec{v}_i
\end{equation}
This means that each value $W_{ij}$ of $\vec{v}_i$ can be calculated as

\begin{multline}
PCorr(\tilde{X_j}, \tilde{X_1})W_{i1} + PCorr(\tilde{X_j}, \tilde{X_2})W_{i2} + \dots \\ 
+ PCorr(\tilde{X_j}, \tilde{X_N})W_{iN} = \lambda_i W_{ij}
\end{multline}
which can be more compactly represent as

\begin{equation}
\sum_l PCorr(\tilde{X}_j, \tilde{X}_l) W_{il} = \lambda_i W_{ij}\label{eq:PCAeigEqProp}
\end{equation}
Since the Pearson correlation coefficient between standardized variables $\tilde{X}_j$ and $\tilde{X}_l$ is given by 

\begin{equation}
PCorr(\tilde{X}_j, \tilde{X}_l) = \frac{1}{Q-1}\sum_k \tilde{X}_{jk}\tilde{X}_{lk}
\end{equation}
Equation~\ref{eq:PCAeigEqProp} can be rewritten as 

\begin{equation}
\frac{1}{Q-1}\sum_l\sum_k \tilde{X}_{jk}\tilde{X}_{lk}W_{il} = \lambda_i W_{ij}\label{eq:corr_eigen_property}
\end{equation}

Using Equation~\ref{eq:corr_eigen_property} in Equation~\ref{eq:corr_pcai_xj}, we obtain

\begin{equation}
PCorr(Y_i, \tilde{X}_j) = \sqrt{\lambda_i} W_{ij}
\end{equation}

Therefore, the Pearson correlation coefficient between PCA component $Y_i$ and standardized variable $\tilde{X}_j$ is given by the square root of the respective eigenvalue multiplied by the $j$th element of the respective eigenvector.

\bibliography{references}

\end{document}